\documentclass[aps,prl,twocolumn,superscriptaddress]{revtex4-1}

\usepackage{hyperref} 
\usepackage{graphicx}
\usepackage{bm}
\usepackage{amsmath}
\usepackage{amssymb}
\usepackage{hyperref}

\hypersetup{pdfstartview={XYZ null null 1.7},
	pdfpagemode=UseNone
	}

\DeclareGraphicsExtensions{.png,.jpg,.eps}
\usepackage{xcolor}

\begin{document}

\author{Wei Chen}
\affiliation{College of Science, Nanjing University of Aeronautics and Astronautics, Nanjing 210016, China}
\affiliation{Institute for Theoretical Physics, ETH Zurich, 8093 Z\"{u}rich, Switzerland}

\author{Hai-Zhou Lu}
\affiliation{Shenzhen Institute for Quantum Science and Engineering and Department of Physics,
Southern University of Science and Technology, Shenzhen 518055, China}
\affiliation{Shenzhen Key Laboratory of Quantum Science and Engineering, Shenzhen 518055, China}

\author{Oded Zilberberg}
\affiliation{Institute for Theoretical Physics, ETH Zurich, 8093 Z\"{u}rich, Switzerland}

\title{Weak Localization and Antilocalization in Nodal-Line Semimetals: Dimensionality and Topological Effects}

\begin{abstract}
New materials such as nodal-line semimetals offer a unique setting for novel transport phenomena. Here, we calculate the quantum correction to conductivity in a disordered nodal-line semimetal. The torus-shaped Fermi surface and encircled $\pi$ Berry flux carried by the nodal loop result in a fascinating interplay between the effective dimensionality of electron diffusion and band topology, which depends on the scattering range of the impurity potential relative to the size of the nodal loop. For a short-range impurity potential, backscattering is dominated by the interference paths that do not encircle the nodal loop, yielding a 3D weak localization effect. In contrast, for a long-range impurity potential, the electrons effectively diffuse in various 2D planes and the backscattering is dominated by the interference paths that encircle the nodal loop. The latter, leads to weak antilocalization with a 2D scaling law. Our results are consistent with symmetry consideration, where the two regimes correspond to the orthogonal and symplectic classes, respectively. Furthermore, we present weak-field magnetoconductivity calculations at low temperatures for realistic experimental parameters, and predict that clear scaling signatures $\propto\sqrt{B}$ and $\propto -\ln B$, respectively. The crossover between the 3D weak localization and 2D weak antilocalization can be probed by tuning the Fermi energy, giving a unique transport signature of the nodal-line semimetal.
\end{abstract}

\date{\today}

\maketitle

\begin{figure}[t!]
 \centering
   \includegraphics[width=\columnwidth]{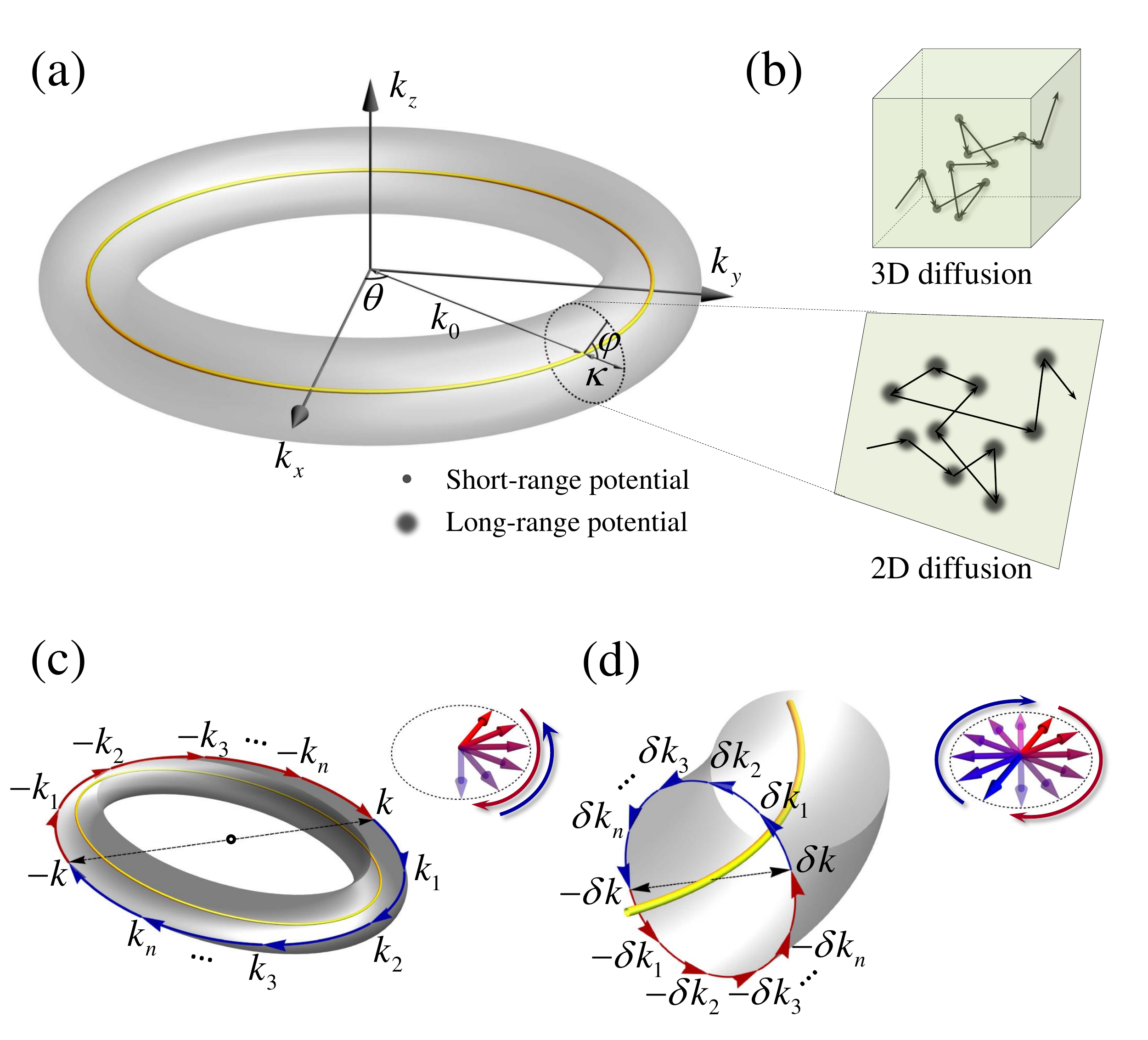}
\caption{Backscattering processes in nodal-line semimetals. (a) Torus-shaped Fermi surface of nodal-line semimetals, with major radius $k_0$, minor radius $\kappa$, toroidal angle $\theta$, and poloidal angle $\varphi$.
(b) The real-space range of the impurity potentials results in 3D and 2D diffusion behaviors, for short-range and long-range impurities, respectively. (c) A coherent backscattering from wavevector $\bm{k}$ to $-\bm{k}$ around the toroidal direction is possible for short-ranged impurity potentials via intermediate states $(\bm{k}_1,\bm{k}_2,...\bm{k}_n)$ and its time-reversed counterpart $(-\bm{k}_n,-\bm{k}_{n-1},...-\bm{k}_1)$. In the circular inset, blue and red arrows depict that the net spinor rotation around the interference path is zero. (d) Backscattering from wavevector $\delta\bm{k}$ to $-\delta\bm{k}$ along the poloidal direction is the dominant process under long-ranged impurity potentials. Here, the net spinor rotation is $2\pi$, contributing a $\pi$ Berry phase to backscaterring, and results in weak antilocalization.
}
\label{fig1}
\end{figure}

The transport properties of materials can be well-approximated by taking into account conduction electrons moving in the presence of a periodic crystalline structure and using Bloch's theory. In the presence of strong disorder, however, this description breaks down and the electrons can localize, leading to Anderson insulators~\cite{Anderson58pr}.
In fact, even weak disorder is sufficient to drive the electronic motion into the so-called quantum diffusive regime, resulting in weak localization (WL)~\cite{Lee85rmp}, which is a precursor of the Anderson localization.
WL is a quantum-mechanical effect where constructive interference between disorder-induced scattering events leads to increased backscattering. The inevitable presence of impurities in materials makes WL effects prominent in experiments, and has become the standard method used for measuring the phase coherence length, as well as its
temperature dependence~\cite{Akkermans07}.

The WL quantum correction depends strongly on the specifics of the electronic system: (i) depending on the dimensionality of the system, the WL correction scales differently with the system size
~\cite{Lee85rmp}; (ii) considering the symmetry class of the system in terms of time-reversal and spin-rotational symmetries, disorder can result in both WL and weak antilocalization (WAL)~\cite{Dyson62jmp,Hikami80ptp}; and (iii) the geometrical and topological properties of the electronic bandstructure~\cite{Kane10rmp,Qi11rmp,Ozawa18arxiv}, can also lead to WAL~\cite{Suzuura02prl,McCann06prl,Lu11prl,Garate12prb,Lu14prl,Lu15prb,Dai16prb}. A paradigmatic example of the latter is graphene, where the $\pi$ Berry phase of the Dirac fermion changes the WL constructive interference into a WAL destructive one~\cite{Ando98jpsj}.
Interestingly, in the study of WL, the Fermi surface of the physical system is commonly taken to be a hypersphere and
the effects of dimensionality and band topology are unrelated.

Relatively new members to the topological phases of matter paradigm are nodal-line semimetals~\cite{Burkov11prb,Kim15prl,Yu15prl,
Heikkila11jetp,Weng15prb,Chen15nl,
Zeng15arxiv,Fang15prb,Mullen15prl,Yamakage16jpsj,
Xie15aplm,Chan16prb,Zhao16prb,Bian16prb,
Bian16nc,Bzdusek16nat,Chenwei17prb,Yan17prb,Ezawa17prb}.
These 3D materials are characterized by bands that
cross along closed loops that
carry a $\pi$ Berry flux~\cite{Burkov11prb}.
A variety of candidates for nodal-line semimetals
have been reported \cite{Kim15prl,Yu15prl,Heikkila11jetp,
Weng15prb,Chen15nl,Zeng15arxiv,Fang15prb,
Yamakage16jpsj,Xie15aplm,Chan16prb,Zhao16prb,
Bian16prb,Bian16nc}, and their experimental
characterization has seen recent progress
using ARPES \cite{Bian16nc,Schoop16nc,
Neupane16prb,Andreas16njp,Takane16prb} and quantum oscillation \cite{Hu16prl,
Hu17prb,Kumar17prb,Pan17arxiv,Li18prl} measurements,
alongside proposals for mesoscopic transport detection schemes~\cite{Chen18prl}.
In most nodal-line semimetals,
the Fermi energy is lifted from the nodal line
thus forming a torus-shaped Fermi surface that encircles the nodal line and its associated $\pi$ Berry flux \cite{Takane18npj}, see Fig. \ref{fig1}(a).


In this Letter, we analyze WL in nodal-line semimetals and find an interesting interplay between the dimensionality of diffusion
and the band topology. We consider two types of disorder with either short-range (SR)
or long-range (LR) impurity potentials relative to the size of the nodal loop.
In the SR limit, the so-called white-noise disorder-induced scattering equally couples all states on the Fermi surface, and WL backscattering
is dominated by interference along the toroidal direction, see Fig. \ref{fig1}(c).
Such an interference loop does not encircle the nodal line and its associated $\pi$ Berry flux. Correspondingly, the electrons diffuse in the full 3D phase-space resulting in 3D WL, similarly to anisotropic conventional metals. The LR limit, can occur due to unconventional screening effects~\cite{Syzranov17prb}. Here, backscattering is dominated by interference along the poloidal direction in reciprocal space that encircles the nodal line, see Fig. \ref{fig1}(d). As a result, we predict that a WAL correction will occur. Importantly, despite the 3D nature of the system, the WAL correction shows a 2D scaling behavior. In the LR scenario, the WAL conductivity correction is proportional to the circumference of the nodal line, or equivalently, to the number of 2D diffusion planes. We discuss possible detection of our prediction using magnetoconductivity experiments, in which tuning the Fermi energy can induce a crossover between the 3D WL and 2D WAL.

Nodal-line semimetals can be generally described by an effective two-band model
\begin{equation}\label{H}
H=\hbar\lambda(k_x^2+k_y^2-k_0^2)\tau_x+\hbar vk_z\tau_y\,,
\end{equation}
where $\tau_{x,y}$ are the Pauli matrices corresponding to the
two-band pseudo-spin space and the two bands cross
at $k_x^2+k_y^2=k_0^2, k_z=0$, and define a nodal loop.
For a Fermi energy that satisfies $E_F\ll\hbar\lambda k_0^2$,
the Hamiltonian \eqref{H} can be
linearized and parametrized to the simple form
\begin{equation}
\mathcal{H}=\hbar v_0\kappa(\cos\varphi\tau_x+\sin\varphi\tau_y)
\end{equation}
through the substitution $k_x=(k_0+\kappa\cos\varphi)\cos\theta,
k_y=(k_0+\kappa\cos\varphi)\sin\theta, k_z=\kappa\sin\varphi/\alpha$,
with $v_0=2\lambda k_0$, and $\alpha=v/v_0$ the ratio between
the velocity along the $z$-direction and the velocity in the $x-y$ plane.
The parameters $\kappa, \theta, \varphi$ are
labeled in Fig. \ref{fig1}(a). For the purposes of this work,
we assume that $E_F$ intersects with the conduction band, and solely include it
in the analysis below. The dispersion of the conduction band is $\varepsilon_{\bm{k}}
=\hbar v_0\kappa$ and its corresponding wavefunction is
$\psi_{\bm{k}}(\bm{r})=[1, e^{i\varphi}]^Te^{i\bm{k}\cdot\bm{r}}/\sqrt{2V}$,
with $V$ the volume of the system. The density of
states at the Fermi energy is $\rho_0=\mathcal{K}E_F/(\alpha h^2v_0^2)$,
which is proportional to the circumference $\mathcal{K}=2\pi k_0$ of the
nodal loop.


The nodal-line
semimetal possesses two types of antiunitary
symmetries $\mathcal{T}_1=K$ and
$\mathcal{T}_2=i\tau_y K$ with $K$ the
complex conjugation, such that
\begin{subequations}\label{symm}
\begin{align}
\mathcal{T}_1\mathcal{H}(\bm{k})\mathcal{T}_1^{-1}&=\mathcal{H}(-\bm{k}), \label{symm1}\\
\mathcal{T}_2\mathcal{H}(\delta\bm{k})\mathcal{T}_2^{-1}&=\mathcal{H}(-\delta\bm{k}), \label{symm2}
\end{align}
\end{subequations}
where the small momentum $\delta\bm{k}=\kappa(\cos\varphi, \sin\varphi/\alpha)$
is defined in a local poloidal plane in momentum space
denoted by the toroidal angle $\theta$, see Fig.~\ref{fig1}(a). Note that
$\mathcal{T}_1$ is the spinless time-reversal symmetry while
$\mathcal{T}_2$ is a local antiunitary symmetry containing a
pseudo-spin inversion in the two-band space.
The latter can be regarded as the spinful time-reversal symmetry
for the 2D subsystem defined by $\theta$, see Fig.~\ref{fig1}(b).
According to symmetry classification, $\mathcal{T}_1$ and $\mathcal{T}_2$ belong to
the orthogonal and symplectic classes,
respectively~\cite{Dyson62jmp,Hikami80ptp}.
Hence, they respectively lead to WL and WAL depending on which physical process dominates
the backscattering [cf. Figs. \ref{fig1}(c) and \ref{fig1}(d)].

The dominant process that leads to
backscattering is determined by the type of disorder, and by the size of the nodal loop.
The disorder potential is expressed by
$
U(\bm{r})=\sum_j \mathcal{U}(\bm{r}-\bm{R}_j),
$
where $\bm{R}_j$ are the positions of the randomly distributed impurities.
Without loss of generality, we set the uniform background of the impurity potential to zero, $\langle U(\bm{r})\rangle_{\text{imp}}=0$, where $\langle\cdots\rangle_{\text{imp}}$ denotes averaging over impurity configurations.
Commonly, the white-noise disorder (SR limit) is considered, i.e., an impurity-potential with constant scattering-strength in reciprocal space~\cite{Lee85rmp}. Here,
we investigate a more general case
where $\langle U(\bm{k})U(\bm{-k}')\rangle_{\text{imp}}=(2\pi)^3 n_i|u_{\bm{k}}|^2 \delta(\bm{k}-\bm{k}')$
with $U(\bm{k})$ $\left[u_{\bm{k}}\right]$ the Fourier component of $U(\bm{r})$ $\left[\mathcal{U}(\bm{r})\right]$, and $n_i$ the concentration of the impurities.
Associating a finite scattering-range to the impurities, $r_{\rm sc}$, results in a confinement of the allowed scattering processes in reciprocal space.
We consider $E_F$ sufficiently low such that $k_0\gg 1/r_{\rm sc} \gg \kappa$, and
therefore the scattering strength between different $\bm{k}$
states can be well characterized solely by $\theta$. For simplicity, we further assume that $u_{\bm{k}}=u(\theta)=u_0f_\Delta(\theta)$ with $f_\Delta(x)=\Theta(x+\Delta)\Theta(-x+\Delta)$, and
$\Theta(x)$ the Heaviside step function. This choice of potential yields the SR limit when $\Delta=\pi$, whereas the LR limit corresponds to $\Delta\rightarrow 0$.

\begin{figure}[t!]
 \centering
                \includegraphics[width=0.9\columnwidth]{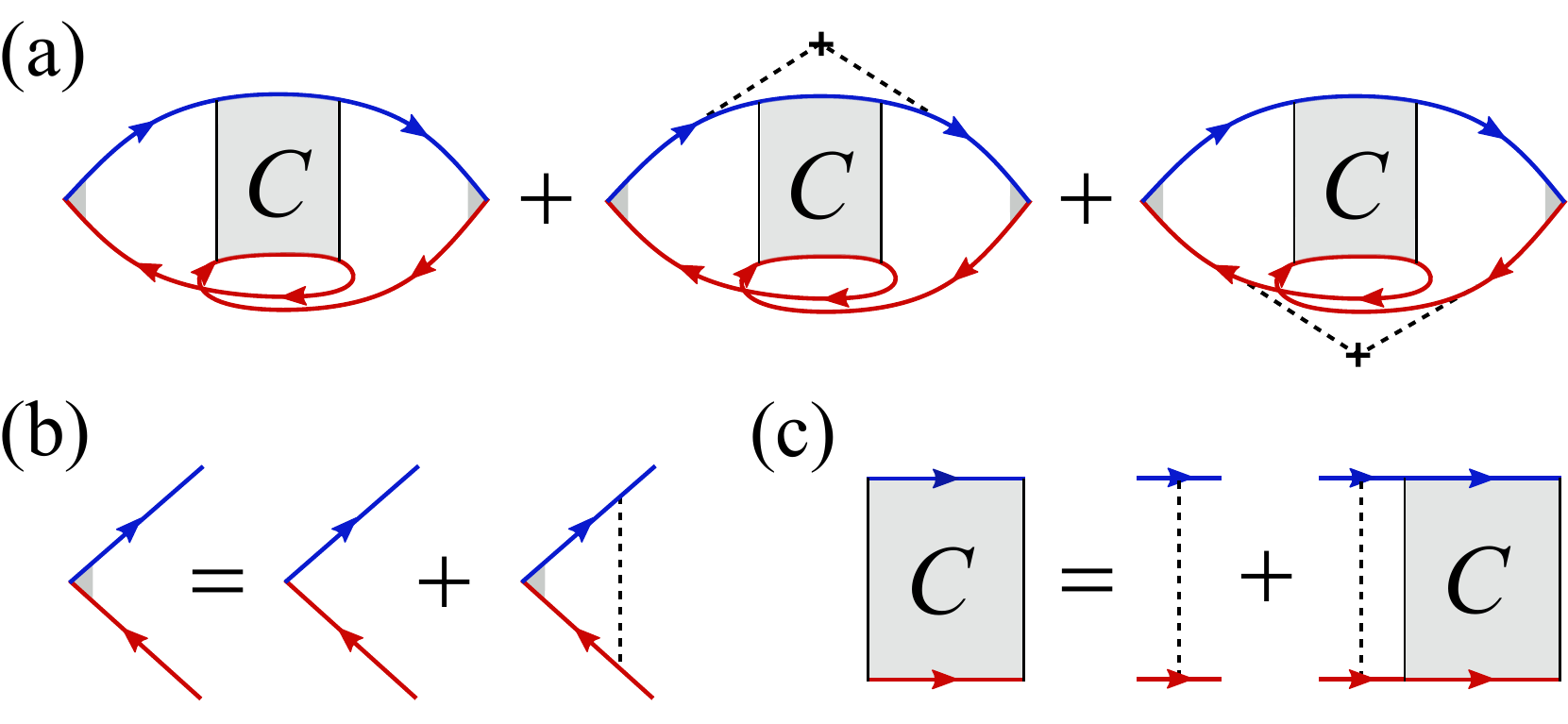}

\caption{Relevant Feynman diagrams. (a) Leading diagrams contributing to the quantum interference correction to conductivity,
containing one bare and two dressed Hikami boxes, cf.~Eq.~\eqref{kubo} and Ref.~\cite{Akkermans07}. The arrowed solid and dashed lines represent Green's functions and impurity scattering, respectively. The gray box marks the Cooperon ladder propagator. (b) Ladder diagram vertex correction to the velocity. (c) Diagrammatic representation of the Bethe-Salpeter equation for the Cooperon correction.
}
\label{fig2}
\end{figure}

We calculate the semiclassical Drude conductivity and obtain a standard 3D metal expression~\cite{supmat}. Furthermore, we derive the quantum interference correction along the $z$-axis
using Feynman diagrams~\cite{generality}, see Fig. \ref{fig2}(a).
In our calculation, we include the three leading-order diagrams, namely, the bare Hikami box \cite{Hikami80ptp},
 and two dressed Hikami boxes~\cite{Chalaev05prb,McCann06prl}. The latter two jointly contribute -1/2 the correction of the former for both the SR and LR limits~\cite{supmat}, similar to the case in graphene \cite{McCann06prl}.
Thus, the overall quantum correction
to the zero-temperature conductivity is~\cite{Akkermans07}
\begin{equation}\label{kubo}
\sigma_z=s\eta_z^2\frac{e^2\hbar}{4\pi V^2}\sum_{\bm{k},\bm{k}'}v^z_{\bm{k}}v^z_{\bm{k}'} G^R_{\bm{k}}G^A_{\bm{k}}G^R_{\bm{k}'}G^A_{\bm{k}'}C_{\bm{k},\bm{k}'},
\end{equation}
where $s=2$ accounts for the spin degeneracy, $\eta_z=2$ is the
factor coming from the vertex
correction to the velocity $v_{\bm{k}}^z$ [cf.~Fig. \ref{fig2}(b) and Ref.~\cite{supmat}]. The impurity-averaged retarded (R) and advanced (A) Green's functions are solved under the first-order Born approximation,
$G^{R,A}_{\bm{k}}(\omega)=1/[\omega-\varepsilon_{\bm{k}}\pm i\hbar/(2\tau_e)]$,
with the elastic scattering time $\tau_e=2\hbar/(n_i\rho_0\tilde{u}^2)$, and
$\tilde{u}^2=\int_0^{2\pi} d\theta|u(\theta)|^2$~\cite{supmat}.
Last, $C_{\bm{k},\bm{k}'}$ is the full Cooperon
satisfying the Bethe-Salpeter equation for the ladder of maximally-crossed diagrams~\cite{Akkermans07}, see Fig.~\ref{fig2}(c).

As a function of the impurity potential different interference loops are available in reciprocal space, see Figs.~\ref{fig1}(c) and \ref{fig1}(d). Correspondingly, the Bethe-Salpeter equation for the Cooperon will be dominated by different processes.
In the SR limit $\Delta=\pi$, the majority of backscattering trajectories come in
pairs via the intermediate states $(\bm{k}_1,\bm{k}_2,...\bm{k}_n)$ and their time-reversed
counterpart $(-\bm{k}_n,-\bm{k}_{n-1},...-\bm{k}_1)$, as imposed by $\mathcal{T}_1$. Importantly, these trajectories do not encircle the nodal line.
The corresponding Bethe-Salpeter equation
is $C_{\bm{k},\bm{k}'}=C^0_{\bm{k},\bm{k}'}
+\frac{1}{V}\sum_{\bm{k}_1}C_{\bm{k},\bm{k}_1}G^R_{\bm{k}_1}
G^A_{\bm{Q}-\bm{k}_1}C^0_{\bm{k}_1,\bm{k}'}$, where the bare
Cooperon is
$C^0_{\bm{k},\bm{k}'}=\langle\tilde{U}_{\bm{k},\bm{k}'
}\tilde{U}_{-\bm{k},-\bm{k}'}\rangle_{\rm imp}/V=\gamma\cos^2\frac{\varphi-\varphi'}{2}$, with $\gamma=n_iu_0^2$,
the momentum $\bm{Q}=\bm{k}+\bm{k}'$,
and $\tilde{U}_{\bm{k},\bm{k}'}=\langle
\psi_{\bm{k}}(\bm{r})|U(\bm{r})|\psi_{\bm{k}'}(\bm{r})\rangle$. Solving the Bethe-Salpeter equations self-consistently and keeping the most divergent contributions for $\bm{Q}\rightarrow0$ yields in the DC-limit ($\omega\rightarrow0$)~\cite{supmat}
\begin{align}
C_{\bm{k},\bm{k}'}&=\frac{\gamma}{2\tau_e}\frac{1}{D_{xy}Q_{xy}^2+D_z{Q}_z^2}, \label{coop1}
\end{align}
where $Q_{xy}=\sqrt{Q_x^2+Q_y^2}$, and $D_{xy}=v_0^2\tau_e/4$, $D_z=\alpha^2 v_0^2\tau_e$ are
diffusion coefficients in the $x-y$ plane and $z$-direction, respectively. We obtain that the Cooperon in the SR limit \eqref{coop1} shows a 3D WL behavior similar to an anisotropic normal metal~\cite{Akkermans07}.

In the LR limit $\Delta\rightarrow0$, the aforementioned backscattering channel is suppressed and the quantum correction is dominated by interference trajectories $(\delta\bm{k}_1,\delta\bm{k}_2,...\delta\bm{k}_n)$
and $(-\delta\bm{k}_n,-\delta\bm{k}_{n-1},...-\delta\bm{k}_1)$, which are
paired by the $\mathcal{T}_2$ symmetry and form a small loop that encircles the nodal
line. In this case, the iterative equation
becomes $C_{\delta\bm{k},\delta\bm{k}'}=C^0_{\delta\bm{k},\delta\bm{k}'}
+\frac{1}{V}\sum_{\delta\bm{k}_1}C_{\delta\bm{k},\delta\bm{k}_1}G^R_{\delta\bm{k}_1
}G^A_{\bm{q}-\delta\bm{k}_1}C^0_{\delta\bm{k}_1,\delta\bm{k}'}$, with
the bare Cooperon $C^0_{\delta\bm{k},\delta\bm{k}'}
=\langle\tilde{U}_{\delta\bm{k},\delta\bm{k}'}\tilde{U}_{-\delta\bm{k},-\delta\bm{k}'}\rangle_{\rm imp}/V
=(\gamma/4)f_\Delta(\theta-\theta')\big[1+2e^{-i(\varphi-\varphi')}+e^{-2i(\varphi-\varphi')}\big]$
and momentum $\bm{q}=\delta\bm{k}+\delta\bm{k}'$ measured from
the nodal line in the plane labeled by $\theta$. Solving the Bethe-Salpeter equations
in this case for $\bm{q}\rightarrow0$ yields in the DC-limit~\cite{supmat}
\begin{align}
C_{\delta\bm{k},\delta\bm{k}'}&=\frac{\gamma}{2\tau_e}\frac{f_\Delta(\theta-\theta') e^{-i(\varphi-\varphi')}}{4D_{xy}q_{xy}^2\cos^2(\theta_q-\theta)+D_z{q}_z^2}, \label{coop2}
\end{align}
where $q_{xy}=\sqrt{q_x^2+q_y^2}$. The scattering here
always occurs accompanied with a spin rotation [see Fig.~\ref{fig1}(d)],
which is the source of the additional geometric phase
$e^{-i(\varphi-\varphi')}$. This geometrical phase leads to the suppression of backscattering
($\varphi-\varphi'=\pi$) and to WAL~\cite{Suzuura02prl}. The result \eqref{coop2} contains two
$\theta$-dependent factors. Hence, for $\Delta\rightarrow0$ and through $f_\Delta(\theta-\theta')$, the scattering occurs between states that lie in the same poloidal plane. Conjointly, the divergent term $\cos^2(\theta_q-\theta)$
vanishes when $\theta_q-\theta=\pi/2$, thus implying that
diffusion cannot happen in the direction perpendicular to
the $\theta$ plane. Combining the above observations, we can reach the intriguing conclusion that in the LR limit electrons exhibit a 2D quantum diffusion [cf.~Fig.~\ref{fig1}(b)].

Inserting the Cooperon expressions into Eq.~\eqref{kubo},
we obtain the correction to the conductivity
in both the SR and LR limits~\cite{supmat}
\begin{subequations}\label{cond}
\begin{align}
\sigma_z^{S}&=-\frac{s\eta_z^2\alpha e^2}{2\pi^2h}(\frac{1}{\ell_e}-\frac{1}{\ell_\phi}),\label{cond1}\\
\sigma_z^{L}&=\frac{s\eta_z^2 \mathcal{K}\alpha e^2 }{2(2\pi)^2 h}\ln(\ell_\phi/\ell_e), \label{cond2}
\end{align}
\end{subequations}
respectively. Here, $\ell_e=v_0\tau_e$ is the mean free path, and
$\ell_\phi$ is the phase coherence length \cite{time}. In the SR limit, a 3D scaling is
obtained for the quantum correction and the overall minus sign indicates WL. In contrast, the LR correction is positive (WAL) and has a 2D scaling law. 
The prefactor $\mathcal{K}$ in $\sigma_z^L$ ensures the unit of
3D conductivity and also enhances the WAL
correction contributed by a large number of 2D diffusion planes. Due to the different dimensionality of the scaling, $\sigma_z^S$ will saturate
as $\ell_\phi\rightarrow\infty$; in contrast, $\sigma_z^L$
always increases as $\ln (\ell_\phi/\ell_e)$ due to the 2D diffusion.

The disorder-induced quantum correction to conductivity
is suppressed by magnetic field induced dephasing~\cite{Akkermans07}. This allows one to experimentally observe the WL and WAL corrections using magnetoconductivity measurements. To calculate the impact of the magnetic field, we impose a quantization
condition to the component of $\bm{Q},\bm{q}$ perpendicular to the
magnetic field, i.e., $Q_{\perp},q_{\perp}=(n+1/2)(4eB/\hbar)\equiv(n+1/2)/\ell_{B}^2$,
where $\ell_{B}$ is the magnetic length. In the SR regime, since the diffusion is 3D, the magnetic field along any direction will lead to dephasing; here we set it to the $z$-direction. For the LR regime,
electrons move in different planes parallel to the $z$-axis. Thus, a magnetic field along the $z$-direction cannot lead to dephasing and we align it along the $x$-direction. We substitute the quantized values of $Q_\perp,q_\perp$
into the Cooperon [Eq.~\eqref{kubo}] and obtain the resulting magnetoconductivity
$\delta\sigma^{S,L}(B)\equiv\sigma_{z}^{S,L}(B)-\sigma^{S,L}_z(0)$, with
\begin{subequations}\label{mag}
\begin{align}
\sigma_{z}^{S}(B)&=-\frac{s\eta_z^2\alpha e^2}{(2\pi)^2 h}\Big[
\Psi(\ell^2_{B}/\ell_e^2+\frac{1}{2})/\ell_e
-\Psi(\ell^2_{B}/\ell_\phi^2+\frac{1}{2})/\ell_\phi \notag \\
&-\int_{1/\ell_\phi}^{1/\ell_e}dx\Psi(\ell_{B}^2x^2+\frac{1}{2})\Big], \label{mag1}\\
\sigma_{z}^{L}(B)&=\frac{s\eta_z^2 \mathcal{K}\alpha e^2}{16\pi^2h}
\int^{2\pi}_0\frac{d\theta}{2\pi}
\Big[\Psi(\frac{\ell_{B}^2}{\ell_e^2\alpha|\sin\theta|}+\frac{1}{2}) \notag\\
&-\Psi(\frac{\ell_{B}^2}{\ell_\phi^2\alpha|\sin\theta|}+\frac{1}{2})\Big], \label{mag2}
\end{align}
\end{subequations}
where $\Psi(x)$ is the digamma function~\cite{supmat}.
In the zero-field limit $B\rightarrow0$,
the results in Eqs.~\eqref{mag} reduce to that of Eqs.~\eqref{cond}.
The average integral over $\theta$ in Eq.~\eqref{mag2}
arises from the fact that electrons in different diffusion planes feel different
magnetic fields perpendicular to the respective plane.

\begin{figure}[t!]
 \centering
                \includegraphics[width=\columnwidth]{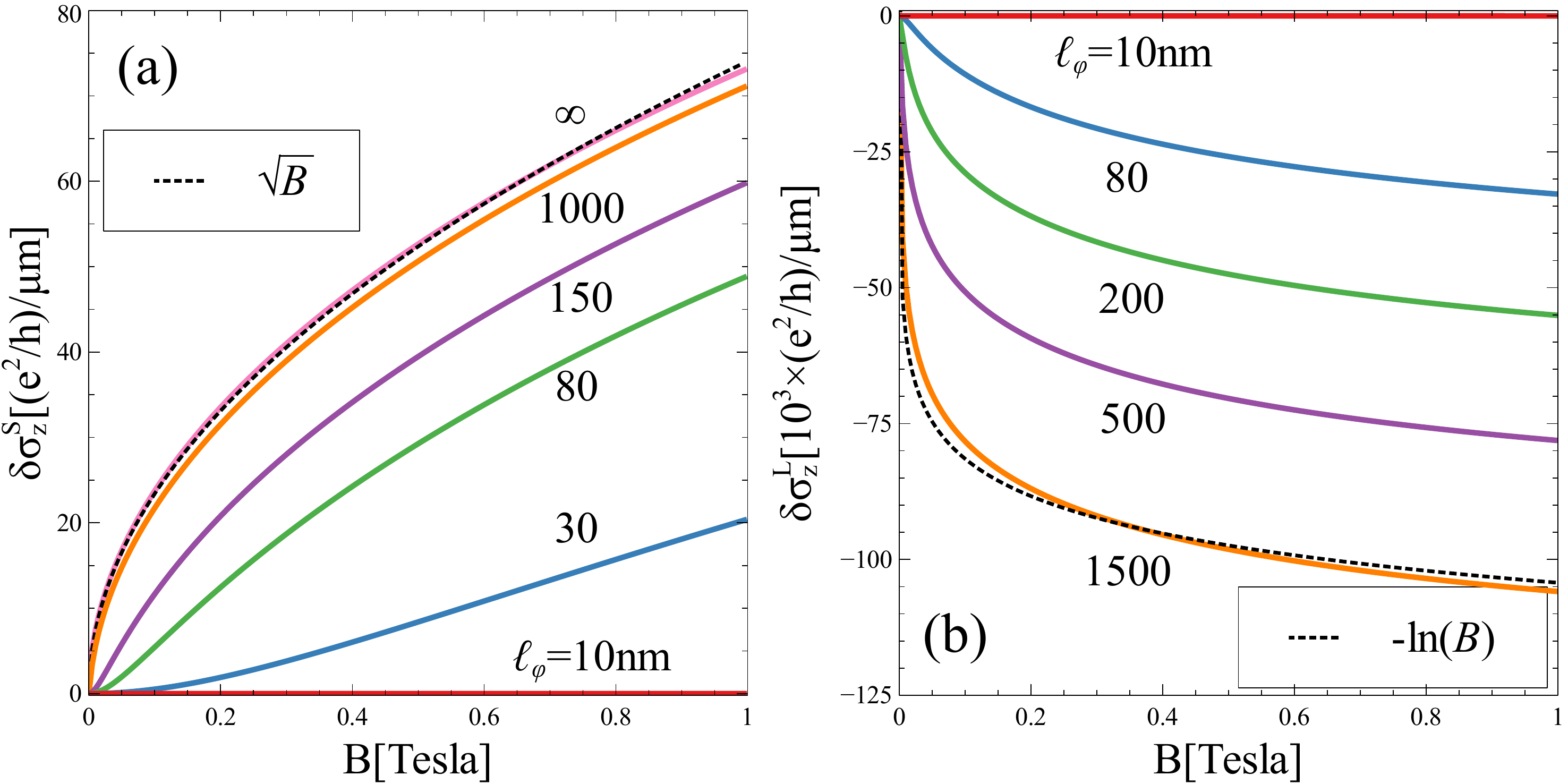}

\caption{The magnetoconductivity $\delta\sigma_z$(B) in the (a) SR limit [cf.~Eq.~\eqref{mag1}] and (b) LR limit [cf.~Eq.~\eqref{mag2}] for different phase coherence lengths $\ell_\phi$. The mean free path is set to $\ell_e=10$nm. The magnetic length is taken to be $\ell_{B}\simeq 12.8 \text{nm}/\sqrt{B}$ with $B$ in Tesla. In (b) $\mathcal{K}=50 \text{nm}^{-1}$ and $\alpha=5$ are used.
}
\label{fig3}
\end{figure}

In Fig.~\ref{fig3}, we plot our magnetoconductivity predictions [Eqs.~\eqref{mag}], where the integrals are evaluated numerically. The WL effect in the SR regime leads to a positive magnetoconductivity [Fig.~\ref{fig3}(a)],
while the WAL effect in the LR regime is revealed by negative magnetoconductivity [Fig.~\ref{fig3}(b)]. We plot the results for various values of $\ell_\phi$, which can be tuned by the temperature of the experiment.
At low temperatures, $\ell_\phi\gg \ell_B$ since
$\ell_\phi\sim 100{\rm nm} - 1\mu{\rm m}$ and $\ell_B\sim 10{\rm nm}$ when $B\sim 0.1-1$T.
The magnetoconductivity in the two scenarios exhibit a different $B$-dependence at low temperatures.
Specifically, $\delta\sigma_z^S(B)\propto\sqrt{B}$
in the SR regime and $\delta\sigma_z^L(B)\propto-\ln B$ in the LR regime,
see fitted dashed lines in Fig.~\ref{fig3}.
Importantly, the $\ln B$ dependence in the LR limit
agrees with our 2D diffusion prediction~\cite{Lee85rmp}, which occurs in a 3D system.
Usually, 2D diffusion results in a much larger WL/WAL effect
than in 3D diffusion. Moreover, the 3D nature of the nodal-line semimetal
contains a large number of 2D effective subsystems, which also significantly enhance the magnetoconductivity~\cite{Cao12prl}.
 Indeed, we observe in Fig.~\ref{fig3} that for reasonable physical parameters,
$\delta\sigma_z^L(B)$ is three orders larger than $\delta\sigma_z^S(B)$, indicating
a very strong signature of WAL in the LR regime.

Our analysis provides a concrete prediction for the impact of weak disorder on transport in nodal-line semimetals.
Current experiments are focused on quantum oscillations in a strong magnetic field
\cite{Hu17prb,Kumar17prb,Pan18sr,Wang16aem,Lv16apl,Ali16sa,Zhang18fp,Li18sb},
and no WL study on nodal-line semimetals has been reported so far.
Our predicted two limits leading to 3D WL and 2D WAL will manifest differently depending on the type of
impurity potentials in the material, and relative to the size of the nodal loop.
In real nodal-line semimetals, the impurity potential is predicted to be of LR-type due to unconventional screening effects~\cite{Syzranov17prb}. Local scattering in
the reciprocal space that leads to WAL is well-defined only when
$\kappa\ll k_0$ [cf.~Fig.~\ref{fig1}(a)]. Otherwise,
if $\kappa$ and $k_0$ are of the same order, the
two kinds of backscattering will coexist (due to toroidal and poloidal interference trajectories),
and hence 3D WL will dominate the transport which stems from the larger
3D phase space for scattering along the toroidal direction.
As a result, a large nodal loop and a low Fermi energy $E_F$ above
the nodal loop is favorable for 2D WAL. As $E_F$
increases, $\kappa$ increases, and WL will overcome WAL.
Furthermore, by increasing $E_F$, the Fermi
surface and the WAL interference loop [Fig.~\ref{fig1}(d)]
may undergo various warping, similar to
the trigonal warping in graphene~\cite{McCann06prl}, thus breaking
the $\mathcal{T}_2$ symmetry, and further suppressing
the 2D WAL. Nevertheless, as long as the warping does not
break time reversal symmetry $\mathcal{T}_1$, the 3D
WL will survive. Consequently, we predict a crossover from 2D WAL to 3D WL to occur
by tuning $E_F$, e.g., by doping or gate tuning on a thin-film sample. This is a unique effect arising from the torus-shaped Fermi surface and the Berry flux in nodal-line semimetals.

Extending our reported analysis to include the effects of spin-orbit coupling,
electron-electron interaction, tilting of the nodal line, and Berry curvature
can lead to a wide-range of interesting results~\cite{supmat}. Furthermore, recent experimental progress
on WAL effect in nodal line semimetal has been reported \cite{An19prb}, which may be
explained by our theory.


\begin{acknowledgments}
We thank J.~L.~Lado, Zhong Wang, Rui Wang and Fengqi Song for helpful discussions.
W.~C. acknowledges the support from the National Natural
Science Foundation of China under Grants No. 11504171, and
the Swiss Government Excellence Scholarship
under the program of China Scholarships Council
(No. 201600160112). O.~Z. thanks the Swiss National Science Foundation for financial support.
H.~L. is supported by the Guangdong Innovative and
Entrepreneurial Research Team Program (2016ZT06D348),
the National Key R \& D Program (2016YFA0301700), the
National Natural Science Foundation of China (11574127),
and the Science, Technology and Innovation Commission of
Shenzhen Municipality (ZDSYS20170303165926217, JCYJ20170412152620376).
\end{acknowledgments}


%

\newpage
\onecolumngrid
\renewcommand{\theequation}{S.\arabic{equation}}
\setcounter{equation}{0}
\renewcommand{\thefigure}{S.\arabic{figure}}
\setcounter{figure}{0}

\section{Supplemental Material for ``Weak Localization and Antilocalization in Nodal Line Semimetals: Dimensionality
and Topological Effects''}

\subsection{Elastic scattering time under first-order Born approximation}

Under the first-order Born approximation, the self-energy is calculated through
\begin{equation}
\Sigma_{\bm{k}}=\frac{1}{V^2}\sum_{\bm{k}_1}G_{0,\bm{k}_1}^R\langle \tilde{U}_{\bm{k},\bm{k}_1}\tilde{U}_{\bm{k}_1,\bm{k}}\rangle_{\text{imp}},
\end{equation}
where $G_{0,\bm{k}_1}^R=1/(\omega-\varepsilon_{\bm{k}_1}+i0^+)$ is the bare retarded Green's function and $\tilde{U}_{\bm{k},\bm{k}_1}=\langle
\psi_{\bm{k}}(\bm{r})|U(\bm{r})|\psi_{\bm{k}_1}(\bm{r})\rangle$.
A direct calculation yields
\begin{equation}
\begin{split}
\Sigma_{\bm{k}}&=\frac{n_i}{V} \sum_{\bm{k}_1}\frac{|u_{\bm{k}-\bm{k}_1}|^2}{\omega-\varepsilon_{\bm{k}_1}+i0^+}\cos^2\frac{\varphi-\varphi_1}{2}\\
&=\frac{n_i}{(2\pi)^3\alpha} \int d\theta_1\int d\varphi_1\int d\kappa_1 \kappa_1(k_0+\kappa_1\cos\varphi_1)\frac{|u_{\bm{k}-\bm{k}_1}|^2}{\omega-\varepsilon_{\bm{k}_1}+i0^+}\cos^2\frac{\varphi-\varphi_1}{2}\\
&\underset{(\kappa_1\ll k_0)}{\simeq} \frac{n_i}{(2\pi)^3\alpha} \int d\theta_1\int d\varphi_1\int d\kappa_1 \kappa_1k_0\frac{|u_{\bm{k}-\bm{k}_1}|^2}{\omega-\varepsilon_{\bm{k}_1}+i0^+}\cos^2\frac{\varphi-\varphi_1}{2}\\
&=\frac{n_i\pi k_0}{(2\pi)^3\alpha} \int d\theta_1|u(\theta-\theta_1)|^2\int d\kappa_1 \kappa_1\frac{1}{\omega-\varepsilon_{\bm{k}_1}+i0^+}\\
&= \frac{n_i\pi}{(2\pi)^2} \int_0^{2\pi} d\theta_1|u(\theta-\theta_1)|^2\int d\varepsilon_1 \frac{\rho(\varepsilon_1)}{\omega-\varepsilon_{1}+i0^+}\\
&= -i\frac{1}{4}n_i\rho(\omega) \tilde{u}^2,
\end{split}
\end{equation}
where $\rho(\omega)$ is the density of states which is set to the value $\rho_0$ at the Fermi energy, $\tilde{u}^2=\int_0^{2\pi} d\theta|u(\theta)|^2$.
With the self-energy, we obtain the full Green's function
\begin{equation}\label{gf}
G^{R,A}_{\bm{k}}(\omega)=\frac{1}{\omega-\varepsilon_{\bm{k}}\pm i\hbar/(2\tau_e)},
\end{equation}
with the elastic scattering time $\tau_e=2\hbar/(n_i\rho_0\tilde{u}^2)$.

\subsection{Vertex correction to velocity}
The group velocities of electrons in the nodal-line semimetal in each of the three directions are
$v^x_{\bm{k}}=v_0\cos\varphi\cos\theta,
v^y_{\bm{k}}=v_0\cos\varphi\sin\theta$, and
$v^z_{\bm{k}}=\alpha v_0\sin\varphi$, where $\theta$ is the toroidal angle, and $\varphi$ the poloidal angle, cf.~Fig.~1 in the main text. Here, we focus on transport along the $z$-direction.
The renormalized velocity $\tilde{v}_{\bm{k}}^z$ can be calculated through an iterative equation that can
be read off from Fig. 2(b) in the main text as
\begin{equation}
\tilde{v}^z_{\bm{k}}=v^z_{\bm{k}}+\frac{1}{V}\sum_{\bm{k}'}
\Gamma^0_{\bm{k},\bm{k}'}G^R_{\bm{k}'}G^A_{\bm{k}'}\tilde{v}^z_{\bm{k}'},
\end{equation}
where $\Gamma^0_{\bm{k},\bm{k}'}=n_i|u(\theta-\theta')|^2\cos^2\frac{\varphi-\varphi'}{2}$ is the bare Diffuson.

We use the trial function $\tilde{v}_{\bm{k}}^z=\eta_z v^z_{\bm{k}}$ to obtain
\begin{equation}
\begin{split}
\frac{\eta_z-1}{\eta_z}\sin\varphi&=\frac{1}{V}\sum_{\bm{k}'}\Gamma^0_{\bm{k},\bm{k}'}G^R_{\bm{k}'}G^A_{\bm{k}'}\sin\varphi'\\
&=\int\frac{d\theta'}{2\pi}\int\frac{d\varphi'}{2\pi}\int d\varepsilon'\rho(\varepsilon')\frac{1}{\omega-\varepsilon'+i\frac{\hbar}{2\tau_e}}\frac{1}{\omega-\varepsilon'
-i\frac{\hbar}{2\tau_e}}n_i|u(\theta-\theta')|^2\cos^2\frac{\varphi-\varphi'}{2}\sin\varphi'\\
&\simeq\rho_0\int\frac{d\theta'}{2\pi}\int\frac{d\varphi'}{2\pi}\int d\varepsilon'\frac{1}{\omega-\varepsilon'+i\frac{\hbar}{2\tau_e}}\frac{1}{\omega-\varepsilon'
-i\frac{\hbar}{2\tau_e}}n_i|u(\theta-\theta')|^2\cos^2\frac{\varphi-\varphi'}{2}\sin\varphi'\\
&=\frac{2\pi n_iu_0^2\tau_e\rho_0}{\hbar}\int\frac{d\theta'}{2\pi}\int\frac{d\varphi'}{2\pi}
f_\Delta(\theta-\theta')\cos^2\frac{\varphi-\varphi'}{2}\sin\varphi'\\
&=\frac{2\Delta n_iu_0^2\tau_e\rho_0}{\hbar}\int\frac{d\varphi'}{2\pi}
\cos^2\frac{\varphi-\varphi'}{2}\sin\varphi'=\frac{1}{2}\sin\varphi\,,
\end{split}
\end{equation}
where we approximated the density of states to be energy-independent in the vicinity of the Fermi energy. Through this analysis, we can conclude that the correction to the velocity in the $z$-direction is $\eta_z=2$, i.e., it is the same as that for graphene. This correction is the same for both the SR and LR limits because $\tau_e$ also depends on $\Delta$.

\subsection{Drude conductivity}
The semiclassical (Drude) conductivity in the $z$-direction can be written as
\begin{equation}
\sigma^{\text{Drude}}_z=s\frac{e^2\hbar}{2\pi V}\sum_{\bm{k}}\tilde{v}_{\bm{k}}^zv_{\bm{k}}^zG^R_{\bm{k}}G^A_{\bm{k}},
\end{equation}
and a direct calculation yields
\begin{equation}
\begin{split}
\sigma^{\text{Drude}}_z&=s\eta_z\frac{e^2\hbar}{2\pi}\alpha^2v_0^2\int\frac{d\theta}{2\pi}\int\frac{d\varphi}{2\pi}\sin^2\varphi
\int\rho(\varepsilon)d\varepsilon\frac{1}{E_F-\varepsilon+i\frac{\hbar}{2\tau_e}}\frac{1}{E_F-\varepsilon-i\frac{\hbar}{2\tau_e}}\\
&=\frac{s\eta_z}{2}e^2\rho_0D_z=2e^2\rho_0D_z,
\end{split}
\end{equation}
where $D_z=\alpha^2 v_0^2\tau_e$. We obtain a similar result to that of 3D metals, and recall that the Drude conductivity does not respond to weak magnetic fields.

\subsection{Cooperon in the SR limit}
In the SR limit, the bare Cooperon reads
\begin{equation}
C^0_{\bm{k},\bm{k}'}
=\frac{1}{V}\langle\tilde{U}_{\bm{k},\bm{k}'}\tilde{U}_{-\bm{k},-\bm{k}'}\rangle_{\rm imp}
=\gamma\cos^2\frac{\varphi-\varphi'}{2},
\end{equation}
with $\gamma=n_iu_0^2$.
The Bethe-Salpeter equation for the full Cooperon $C_{\bm{k},\bm{k}'}$ is [cf.~Fig. 2(c) in the main text]
\begin{equation}
C_{\bm{k},\bm{k}'}=C^0_{\bm{k},\bm{k}'}
+\frac{1}{V}\sum_{\bm{k}_1}C_{\bm{k},\bm{k}_1}G^R_{\bm{k}_1}G^A_{\bm{Q}-\bm{k}_1}C^0_{\bm{k}_1,\bm{k}'},
\end{equation}
where $\bm{Q}=\bm{k}+\bm{k}'$.
Transferring the summation over $\bm{k}_1$ into an integral over $(\varepsilon_1,\theta_1,\varphi_1)$ yields
\begin{equation}
\begin{split}
C_{\bm{k},\bm{k}'}
&=C^0_{\bm{k},\bm{k}'}
+\rho_0\int\frac{d\theta_1}{2\pi}\int\frac{d\varphi_1}{2\pi} \int d\varepsilon_1 C_{\bm{k},\bm{k}_1}G^R_{\bm{k}_1}G^A_{\bm{Q}-\bm{k}_1}C^0_{\bm{k}_1,\bm{k}'}\\
&=C^0_{\bm{k},\bm{k}'}
+\rho_0\int\frac{d\theta_1}{2\pi}\int\frac{d\varphi_1}{2\pi} C_{\bm{k},\bm{k}_1}C^0_{\bm{k}_1,\bm{k}'}\int d\varepsilon_1 G^R_{\bm{k}_1+\bm{Q}}G^A_{\bm{k}_1}\\
&=C^0_{\bm{k},\bm{k}'}
+\frac{2\pi\tau_e\rho_0}{\hbar}\int\frac{d\theta_1}{2\pi}\int\frac{d\varphi_1}{2\pi} C_{\bm{k},\bm{k}_1}C^0_{\bm{k}_1,\bm{k}'}\frac{1}{1-i\omega\tau_e/\hbar+i\bm{v}_1\cdot\bm{Q}\tau_e}.\\
\end{split}
\label{cooperon1}
\end{equation}

In the diffusive limit, $Q\ell_e\ll1, \omega\tau_e\ll1$, We can expand Eq.~\eqref{cooperon1}
\begin{equation}
C_{\bm{k},\bm{k}'}\simeq C^0_{\bm{k},\bm{k}'}
+\frac{2\pi\tau_e\rho_0}{\hbar}\int\frac{d\theta_1}{2\pi}\int\frac{d\varphi_1}{2\pi} C_{\bm{k},\bm{k}_1}C^0_{\bm{k}_1,\bm{k}'}\big[(1+i\omega\tau_e/\hbar-\omega^2\tau_e^2/\hbar^2)
+(2\omega\tau_e^2/\hbar-i\tau_e)\bm{v}_1\cdot\bm{Q}-\tau_e^2(\bm{v}_1\cdot\bm{Q})^2\big]\,,
\end{equation}
and rewrite the resulting expression in the following form
\begin{equation}\label{bs1}
C_{\bm{k},\bm{k}'}=C^0_{\bm{k},\bm{k}'}
+\frac{2\pi\tau_e\rho_0}{\hbar}\int\frac{d\theta_1}{2\pi}\int\frac{d\varphi_1}{2\pi} C_{\bm{k},\bm{k}_1}C^0_{\bm{k}_1,\bm{k}'}\chi(\theta_1,\varphi_1,\bm{Q})\,,
\end{equation}
with
\begin{align}
\chi(\theta_1,\varphi_1,\bm{Q})
=&f_1+f_2\Big(\cos\varphi_1\cos(\theta_1-\theta_Q) Q_{xy}+\alpha\sin\varphi_1 Q_z\Big)\\
&+f_3\Big(\cos\varphi_1\cos(\theta_1-\theta_Q) Q_{xy}+\alpha\sin\varphi_1 Q_z\Big)^2\,,\nonumber
\end{align}
and
\begin{equation}
\begin{split}
f_1=1+i\omega\tau_e/\hbar-\omega^2\tau_e^2/\hbar^2;\ \ f_2=(2\omega\tau_e^2/\hbar-i\tau_e) v_0;\ \
f_3=-\tau_e^2v_0^2;\\
\theta_Q=\tan^{-1}(Q_y/Q_x);\ \ Q_{xy}=\sqrt{Q_x^2+Q_y^2}\,,
\end{split}
\end{equation}
where $v_0$ the magnitude, $\theta_1$ is the toroidal angle, and $\varphi_1$ the poloidal angle of $\bm{v}_1$.

Considering that the bare Cooperon depends solely on $\varphi$ and $\varphi'$,
we postulate that the general form of the full Cooperon can be expressed as
$C_{\bm{k},\bm{k}'}=\gamma\zeta(\varphi,\varphi')$ with
\begin{equation}
\begin{split}
\zeta(\varphi,\varphi')&=a_0+a_1\cos\varphi+a_2\cos\varphi'+a_3\sin\varphi+a_4\sin\varphi'\\
&+a_5\cos\varphi\cos\varphi'+a_6\sin\varphi\sin\varphi'+a_7\cos\varphi\sin\varphi'+a_8\sin\varphi\cos\varphi'\,,
\end{split}
\end{equation}
and $a_i$ being arbitrary parameters.

Inserting the general form of $C_{\bm{k},\bm{k}'}$
into Eq. \eqref{bs1}, the Cooperon's iterative equation reduces to
\begin{equation}
\begin{split}
\zeta(\varphi,\varphi')=\cos^2\frac{\varphi-\varphi'}{2}
+2\int\frac{d\theta_1}{2\pi}\int\frac{d\varphi_1}{2\pi} \zeta(\varphi,\varphi_1)\cos^2\frac{\varphi_1-\varphi'}{2}\chi(\theta_1,\varphi_1,\bm{Q})\,,
\end{split}
\label{selfconstSR}
\end{equation}
where in the SR limit we have used $\tau_e\rho_0=\hbar/(\pi \gamma)$. Solving Eq.~\eqref{selfconstSR}, we obtain expressions for all of the $a_i$ coefficients, where the most
divergent contribution in the limit $\bm{Q}\rightarrow0, \omega\rightarrow0$ is
\begin{equation}
a_0=\frac{1}{2}\frac{1}{v_0^2 Q_{xy}^2\tau_e^2/4+\alpha^2 v_0^2Q_z^2\tau_e^2}\,,
\end{equation}
and we obtain the full Cooperon in the SR limit [cf.~Eq.~(5) in the main text]
\begin{equation}\label{wnc}
C_{\bm{k},\bm{k}'}=\frac{\gamma}{2\tau_e}\frac{1}{D_{xy}Q_{xy}^2+D_zQ_z^2}\,,
\end{equation}
with $D_{xy}=v_0^2\tau_e/4$.

\subsection{Cooperon in the LR limit}

In the LR limit, $\Delta\rightarrow0$, the backscattering
from $\bm{k}$ to $-\bm{k}$ cannot be achieved. Instead, we consider
the interference loop along the poloidal direction of the Fermi surface,
see Fig. 1(d) in the main text. We introduce the local momentum
$\delta\bm{k}=\kappa(\cos\varphi,\sin\varphi/\alpha)$ defined in a local plane
in the reciprocal space denoted by $\theta$.
The in-plane backscattering is $\delta\bm{k}\rightarrow-\delta\bm{k}$, or equivalently,
$\theta\rightarrow\theta, \varphi\rightarrow\varphi+\pi$.
The bare Cooperon is
\begin{equation}
\begin{split}
C^0_{\delta\bm{k},\delta\bm{k}'}&=\frac{1}{V}\langle\tilde{U}_{\delta\bm{k},\delta\bm{k}'}\tilde{U}_{-\delta\bm{k},-\delta\bm{k}'}\rangle_{\rm imp}
=\gamma f_\Delta(\theta-\theta')\frac{1}{4}\big[1+2e^{-i(\varphi-\varphi')}+e^{-2i(\varphi-\varphi')}\big]\,.
\end{split}
\end{equation}
The Bethe-Salpeter equation for the full Cooperon $C_{\delta\bm{k},\delta\bm{k}'}$ is
\begin{equation}
C_{\delta\bm{k},\delta\bm{k}'}=C^0_{\delta\bm{k},\delta\bm{k}'}
+\frac{1}{V}\sum_{\delta\bm{k}_1}C_{\delta\bm{k},\delta\bm{k}_1}G^R_{\delta\bm{k}_1
}G^A_{\bm{q}-\delta\bm{k}_1}C^0_{\delta\bm{k}_1,\delta\bm{k}'}\,,
\end{equation}
with $\bm{q}=\delta\bm{k}+\delta\bm{k}'$. Note that the summation over
$\delta\bm{k}_1$ still means the summation over all $\bm{k}_1$ states, since
all states in the 3D reciprocal space contribute to the Cooperon. Performing the integral
over $\varepsilon_1$ we obtain
\begin{equation}\label{bs2}
\begin{split}
C_{\delta\bm{k},\delta\bm{k}'}&=C^0_{\delta\bm{k},\delta\bm{k}'}
+\frac{2\pi\tau_e\rho_0}{\hbar}\int\frac{d\theta_1}{2\pi}\int\frac{d\varphi_1}{2\pi} C_{\delta\bm{k},\delta\bm{k}_1}C^0_{\delta\bm{k}_1,\delta\bm{k}'}\chi(\theta_1,\varphi_1,\bm{q})\\
\chi(\theta_1,\varphi_1,\bm{q})
&=f_1+f_2\Big(\cos\varphi_1\cos(\theta_1-\theta_q) q_{xy}+\alpha\sin\varphi_1 q_z\Big)\\
&\ \ \ +f_3\Big(\cos\varphi_1\cos(\theta_1-\theta_q) q_{xy}+\alpha\sin\varphi_1 q_z\Big)^2\\
\theta_q&=\tan^{-1}(q_y/q_x), q_{xy}=\sqrt{q_x^2+q_y^2}\,,
\end{split}
\end{equation}
which resembles the SR-limit result [Eq.~\eqref{bs1}].

Now, the paired interference paths are local in the reciprocal space, lying
in various 2D planes. The bare Cooperon contains the term
$f_\Delta(\theta-\theta')$, confining the initial and final states
to the phase space $|\theta-\theta'|<\Delta$.
In the limit $\Delta\rightarrow0$, it is reasonable to
assume that for the full Cooperon,
the scattering also occurs within the same phase space. So we postulate that the full Cooperon takes the general form of $C_{\delta\bm{k},\delta\bm{k}'}=\gamma f_\Delta(\theta-\theta')\xi(\varphi,\varphi')$, with
\begin{equation}
\begin{split}
\xi(\varphi,\varphi')&=b_0+b_1e^{-i\varphi}+b_2e^{i\varphi'}+b_3e^{-2i\varphi}+b_4e^{2i\varphi'}\\
&+b_5e^{-i(\varphi-\varphi')}
+b_6e^{-i\varphi+2i\varphi'}+b_7e^{-2i\varphi+i\varphi'}+b_8e^{-2i(\varphi-\varphi')}\,,
\end{split}
\end{equation}
with $b_i$ the parameters. Inserting the general form
$C_{\delta\bm{k},\delta\bm{k}'}$ into Eq. \eqref{bs2},
and making the approximation $f_\Delta(\theta-\theta_1)f_\Delta(\theta_1-\theta')
\simeq f_\Delta(\theta-\theta')f_\Delta(\theta_1-\theta)$ for $\Delta\rightarrow0$, the
iterative equation for $\theta=0$ reduces to
\begin{equation}
\begin{split}
\xi(\varphi,\varphi')&=\frac{1}{4}\big[1+2e^{-i(\varphi-\varphi')}+e^{-2i(\varphi-\varphi')}\big]\\
&+\frac{2\pi}{\Delta}
\int_{-\Delta}^{\Delta}\frac{d\theta_1}{2\pi}\int\frac{d\varphi_1}{2\pi}\xi(\varphi,\varphi_1)
\frac{1}{4}\big[1+2e^{-i(\varphi_1-\varphi')}+e^{-2i(\varphi_1-\varphi')}\big]\chi(\theta_1,\varphi_1,\bm{q}),
\end{split}
\end{equation}
where we have used $\tau_e\rho_0=\hbar/(\Delta \gamma)$ in the LR limit.
Solving the equation above, we obtain the most
divergent contribution in the limit $q\rightarrow0, \omega\rightarrow0$, which is
\begin{equation}
b_5=\frac{1}{2}\frac{1}{v_0^2q_{xy}^2\tau_e^2\cos^2\theta_q+\alpha^2v_0^2q_z^2\tau_e^2}.
\end{equation}
For the Cooperon with a general $\theta$,
we replace $\theta_q$ by $\theta_q-\theta$, yielding
\begin{equation}\label{coop2}
C_{\delta\bm{k},\delta\bm{k}'}=\frac{\gamma}{2\tau_e}\frac{f_\Delta(\theta-\theta') e^{-i(\varphi-\varphi')}}{4D_{xy}q_{xy}^2\cos^2(\theta_q-\theta)+D_zq_z^2},
\end{equation}
leading to Eq. (6) in the main text.

\subsection{Conductivity correction in the SR limit}
The quantum interference correction to the conductivity
can be calculated by the Feynman diagrams in Fig. 2(a)
in the main text, i.e., the overall quantum correction
is the sum of three Hikami-box terms
\begin{equation}
\sigma_z^S=\sigma_{1z}^S+2\sigma_{2z}^S,
\end{equation}
where the first term is referred to as ``bare Hikami box''
and the second term constitutes two ``dressed Hikami boxes'', which contribute equally.
In the following, we will calculate them and show that $\sigma_{2z}^S=-(1/4)\sigma_{1z}^S$.
As a result, the final correction to the conductivity is
\begin{equation}
\sigma_z^S=\sigma_{1z}^S/2.
\end{equation}

\subsubsection{Bare Hikami box in the SR limit}
The quantum interference correction to the conductivity by the bare Hikami box can be calculated through
\begin{equation}
\begin{split}
\sigma_{1z}^S&=s\eta_z^2\frac{e^2\hbar}{2\pi V^2}\sum_{\bm{k},\bm{k}'}v^z_{\bm{k}}v^z_{\bm{k}'} G^R_{\bm{k}}G^A_{\bm{k}}G^R_{\bm{k}'}G^A_{\bm{k}'}C_{\bm{k},\bm{k}'}\\
&=s\eta_z^2\frac{e^2\hbar}{2\pi V^2}\sum_{\bm{k},\bm{Q}}v^z_{\bm{k}}v^z_{\bm{Q}-\bm{k}}G^R_{\bm{k}}G^A_{\bm{k}}G^R_{\bm{Q}-\bm{k}}G^A_{\bm{Q}-\bm{k}}C_{\bm{k},\bm{Q}-\bm{k}}\,.
\end{split}
\end{equation}
The Cooperon $C_{\bm{k},\bm{Q}-\bm{k}}$ depends solely on $\bm{Q}$ [cf.~Eq.~(5) in the main text], and it diverges at $\bm{Q}=0$, so that the above expression can be simplified to
\begin{equation}
\begin{split}
\sigma_{1z}^S&=s\eta_z^2\frac{e^2\hbar}{2\pi V^2}\sum_{\bm{Q}}C_{\bm{k},\bm{Q}-\bm{k}}\sum_{\bm{k}}v^z_{\bm{k}}v^z_{-\bm{k}}G^R_{\bm{k}}G^A_{\bm{k}}G^R_{-\bm{k}}G^A_{-\bm{k}}\,.
\end{split}
\end{equation}
We first calculate the second summation
\begin{equation}\label{sumG}
\begin{split}
\sum_{\bm{k}}v^z_{\bm{k}}v^z_{-\bm{k}}G^R_{\bm{k}}G^A_{\bm{k}}G^R_{-\bm{k}}G^A_{-\bm{k}}
&=V\rho_0\int^{2\pi}_0\frac{d\theta}{2\pi}\int_0^{2\pi}\frac{d\varphi}{2\pi}\int d\varepsilon v^z_{\bm{k}}v^z_{-\bm{k}}G^R_{\bm{k}}G^A_{\bm{k}}G^R_{-\bm{k}}G^A_{-\bm{k}}\\
&=-(\alpha v_0)^2V\rho_0\int^{2\pi}_0\frac{d\theta}{2\pi}\int_0^{2\pi}\frac{d\varphi}{2\pi} \sin^2\varphi\int d\varepsilon G^R_{\bm{k}}G^A_{\bm{k}}G^R_{-\bm{k}}G^A_{-\bm{k}}\\
&=-\frac{(\alpha v_0)^2V\rho_0}{2}\int d\varepsilon G^R_{\bm{k}}G^A_{\bm{k}}G^R_{-\bm{k}}G^A_{-\bm{k}}\\
&=-\frac{(\alpha v_0)^2V\rho_0}{2}\int d\varepsilon \frac{1}{(E_F-\varepsilon+i\frac{\hbar}{2\tau_e})^2}\frac{1}{(E_F-\omega-\varepsilon-i\frac{\hbar}{2\tau_e})^2}\\
&\underset{(\omega\rightarrow0)}{=}-2\pi(\alpha v_0)^2V\rho_0\tau_e^3/\hbar^3\,.
\end{split}
\end{equation}
Thus, the quantum interference correction to the conductivity reduces to
\begin{equation}\label{condwn}
\begin{split}
\sigma_{1z}^S&=-s\eta_z^2\frac{e^2}{V\hbar^2}(\alpha v_0)^2\rho_0\tau_e^3\sum_{\bm{Q}}\frac{\gamma}{2}\frac{1}{v_0^2 Q_{xy}^2\tau_e^2/4+\alpha^2 v_0^2Q_z^2\tau_e^2}\\
&=-s\eta_z^2\frac{e^2}{V\hbar^2}(\alpha v_0)^2\rho_0\tau_e^3\frac{2\gamma}{v_0^2\tau_e^2}\sum_{\bm{Q}}\frac{1}{Q_{xy}^2+4\alpha^2Q_z^2}\\
&=-\frac{4s\eta_z^2\alpha^2 e^2}{h}\frac{1}{V}\sum_{\bm{Q}}\frac{1}{Q_{xy}^2+4\alpha^2Q_z^2}\\
&=-\frac{4s\eta_z^2\alpha^2 e^2}{h}\frac{1}{(2\pi)^3}\int d^3Q\frac{1}{Q_x^2+Q_y^2+4\alpha^2Q_z^2}\\
&\underset{(Q'_z=2\alpha Q_z)}{=}-\frac{2s\eta_z^2\alpha e^2}{h}\frac{1}{(2\pi)^3}\int dQ_xdQ_ydQ'_z\frac{1}{Q_x^2+Q_y^2+Q{'}_z^2}\\
&=-\frac{2s\eta_z^2\alpha e^2}{h}\frac{1}{2\pi^2}\int_{1/\ell_\phi}^{1/\ell_e} \frac{Q'^2dQ'}{Q'^2}\,,
\end{split}
\end{equation}
where $Q'^2=Q_x^2+Q_y^2+{Q'_z}^2$, and the cutoff of the integral is defined in terms of the mean free path $\ell_e$ and the phase coherence length $\ell_\phi$. Finally, we obtain the correction to the conductivity by the bare Hikami box in the SR limit
\begin{equation}
\sigma_{1z}^S=-\frac{s\eta_z^2\alpha e^2}{\pi^2h}(\frac{1}{\ell_e}-\frac{1}{\ell_\phi}).
\end{equation}

\subsubsection{Dressed Hikami box in the SR limit}
The dressed Hikami box correction includes two diagrams in Fig. 2(a) in the main text. They contribute equally to the conductivity. In the following, we calculate the retarded Hikami box [second diagram in Fig.~2(a) in the main text]
\begin{equation}
\begin{split}
\sigma_{2z}^S&=s\eta_z^2\frac{e^2\hbar}{2\pi V^3}\sum_{\bm{k},\bm{Q},\bm{k}_1}v^z_{\bm{k}}v^z_{\bm{Q}-\bm{k}_1}
G^A_{\bm{k}}G^A_{\bm{Q}-\bm{k}_1}C_{\bm{k}_1,\bm{Q}-\bm{k}}
G^R_{\bm{k}}G^R_{\bm{k}_1}
C^0_{\bm{k},\bm{k}_1}G^R_{\bm{Q}-\bm{k}}G^R_{\bm{Q}-\bm{k}_1}\,.
\end{split}
\end{equation}
Inserting the bare and full Cooperon expressions into the
expression above and using its divergence at $\bm{Q}\rightarrow0$ yields
\begin{equation}
\begin{split}
\sigma_{2z}^S&=s\eta_z^2\frac{e^2\hbar}{2\pi V^3}\sum_{\bm{Q}}C_{\bm{k}_1,\bm{Q}-\bm{k}}\sum_{\bm{k},\bm{k}_1}v^z_{\bm{k}}v^z_{-\bm{k}_1}
G^A_{\bm{k}}G^A_{-\bm{k}_1}
G^R_{\bm{k}}G^R_{\bm{k}_1}
C^0_{\bm{k},\bm{k}_1}G^R_{-\bm{k}}G^R_{-\bm{k}_1}\\
&=s\eta_z^2\frac{e^2\hbar}{2\pi V}\Big\{\sum_{\bm{Q}}C_{\bm{k}_1,\bm{Q}-\bm{k}}\int \frac{d\theta}{2\pi}\int \frac{d\varphi}{2\pi}\int \rho(\varepsilon)d\varepsilon \int \frac{d\theta_1}{2\pi}\int \frac{d\varphi_1}{2\pi}\int \rho(\varepsilon_1)d\varepsilon_1 \\
&\ \ \ \times v^z_{\bm{k}}v^z_{-\bm{k}_1}
G^A_{\bm{k}}G^A_{-\bm{k}_1}
G^R_{\bm{k}}G^R_{\bm{k}_1}
C^0_{\bm{k},\bm{k}_1}G^R_{-\bm{k}}G^R_{-\bm{k}_1}\Big\}\\
&=s\eta_z^2\frac{e^2\hbar}{2\pi V}(-\alpha^2\rho_0^2v_0^2\gamma)\frac{1}{8}\big(\int d\varepsilon G^A_{\bm{k}}G^R_{\bm{k}}G^R_{-\bm{k}}\big)^2\sum_{\bm{Q}}C_{\bm{k}_1,\bm{Q}-\bm{k}}\\
&=\frac{2\pi s\eta_z^2}{8V}e^2\hbar\alpha^2\rho_0^2v_0^2\gamma\tau_e^4\sum_{\bm{Q}}C_{\bm{k}_1,\bm{Q}-\bm{k}}\\
\end{split}
\end{equation}
Inserting the Cooperon we obtain
\begin{equation}
\sigma_{2z}^S=\frac{s\eta_z^2\alpha^2 e^2}{h}\frac{1}{V}\sum_{\bm{Q}}\frac{1}{Q_{xy}^2+4\alpha^2Q_z^2}=-\frac{1}{4}\sigma^S_{2z}.
\end{equation}
Therefore, the correction due to dressed Hikami boxes
is $-1/4$ of the bare one. A similar calculation for the advanced Hikami box [third diagram in Fig.~2(a) in the main text] reveals that it is generating an equivalent contribution as the retarded one. As a result, the final quantum correction to the conductivity in the SR limit is
\begin{equation}
\sigma_z^S=\frac{1}{2}\sigma_{1z}^S=-\frac{s\eta_z^2\alpha e^2}{2\pi^2h}(\frac{1}{\ell_e}-\frac{1}{\ell_\phi})\,,
\end{equation}
which is Eq.~(7a) in the main text.

\subsection{Conductivity correction in the LR limit}
Similar to the SR limit, the quantum interference correction to the conductivity
in the LR limit also contains two terms, one bare Hikami box and two dressed Hikami boxes, and the overall result is
\begin{equation}
\sigma_z^L=\sigma_{1z}^L+2\sigma_{2z}^L\,.
\end{equation}
Interestingly, in the following we will find that $\sigma_{2z}^L=-(1/4)\sigma_{1z}^L$, the same relation between two terms as that in the SR limit.
As a result, the final correction to the conductivity is
\begin{equation}
\sigma_z^L=\sigma_{1z}^L/2\,.
\end{equation}

\subsubsection{Bare Hikami box in the LR limit}
In the LR limit, the quantum interference correction to the conductivity by the bare Hikami box is
\begin{equation}
\sigma_{1z}^L=s\eta_z^2\frac{e^2\hbar}{2\pi V^2}\sum_{\bm{k},\bm{k}'}v^z_{\bm{k}}v^z_{\bm{k}'} G^R_{\bm{k}}G^A_{\bm{k}}G^R_{\bm{k}'}G^A_{\bm{k}'}C_{\delta\bm{k},\delta\bm{k}'}\,,
\end{equation}
where we have inserted the Cooperon $C_{\delta\bm{k},\delta\bm{k}'}$ in the LR limit, meaning the
interference paths are paired in the poloidal direction of the Fermi surface.
Note that the summation is still over $\bm{k},\bm{k}'$ in the full 3D reciprocal space.
We change the labels of the momentum $\bm{k}, \bm{k}'$ to $\delta\bm{k}, \delta\bm{k}'$,
and the scattering is confined by $f_\Delta(\theta-\theta')$ to
the phase space $|\theta-\theta'|<\Delta$. Inserting the Cooperon formula we obtain
\begin{equation}
\begin{split}
\sigma_{1z}^L&=s\eta_z^2\frac{e^2\hbar}{2\pi V^2}\sum_{\bm{k}}\frac{V}{(2\pi)^3}\int d^3k' v^z_{\delta\bm{k}}v^z_{\delta\bm{k}'} G^R_{\delta\bm{k}}G^A_{\delta\bm{k}}G^R_{\delta\bm{k}'}G^A_{\delta\bm{k}'}\\
&\times \frac{f_\Delta(\theta-\theta')}{2}\frac{\gamma e^{-i(\varphi-\varphi')}}{v_0^2q_{xy}^2\tau_e^2\cos^2(\theta_q-\theta)+\alpha^2v_0^2q_z^2\tau_e^2}\,.
\end{split}
\end{equation}
For the backscattering, $\varphi-\varphi'=\pi$, the phase factor contributes $-1$. The function $f_\Delta(\theta-\theta')$
confines the scattering states $\delta\bm{k},\delta\bm{k}'$ in the same plane, so $\theta_q\simeq\theta$.
Moreover, the Cooperon diverges as $\bm{q}=\delta\bm{k}+\delta\bm{k}'\rightarrow0$.
Using these conditions, the above expression further reduces to
\begin{equation}
\begin{split}
\sigma_{1z}^L&=s\eta_z^2\frac{e^2\hbar}{2\pi V^2}\sum_{\bm{k}}\frac{V}{(2\pi)^3}\int d\theta'\int d\varphi'\int d\kappa'\kappa'\frac{k_0}{\alpha} v^z_{\delta\bm{k}}v^z_{-\delta\bm{k}} G^R_{\delta\bm{k}}G^A_{\delta\bm{k}}G^R_{-\delta\bm{k}}G^A_{-\delta\bm{k}}\\
&\times \frac{f_\Delta(\theta-\theta')}{2}\frac{-\gamma }{v_0^2q_{xy}^2\tau_e^2\cos^2(\theta_q-\theta)+\alpha^2v_0^2q_z^2\tau_e^2}\\
&=s\eta_z^2\frac{e^2\hbar}{2\pi V^2}\frac{k_0\Delta}{\alpha}\sum_{\bm{k}}\frac{V}{(2\pi)^3} v^z_{\delta\bm{k}}v^z_{-\delta\bm{k}} G^R_{\delta\bm{k}}G^A_{\delta\bm{k}}G^R_{-\delta\bm{k}}G^A_{-\delta\bm{k}}\\
&\times \int d\varphi'\int d\kappa'\kappa' \frac{-\gamma }
{v_0^2q_{xy}^2\tau_e^2+\alpha^2v_0^2q_z^2\tau_e^2}\,.
\end{split}
\end{equation}
Using $G_{-\delta\bm{k}}=G_{\delta\bm{k}}=G_{\bm{k}}$ and restoring the momentum labeling by $\bm{k}$ we obtain
\begin{equation}
\begin{split}\label{cond2}
\sigma_{1z}^L&=s\eta_z^2\frac{e^2\hbar}{2\pi V^2}\frac{k_0\Delta}{\alpha}\sum_{\bm{k}} -{v^z_{\bm{k}}}^2 {G^R_{\bm{k}}}^2{G^A_{\bm{k}}}^2\\
&\times \frac{V}{(2\pi)^3}\int d\varphi'\int d\kappa'\kappa' \frac{-\gamma }{v_0^2q_{xy}^2\tau_e^2+\alpha^2v_0^2q_z^2\tau_e^2}\,.
\end{split}
\end{equation}
The summation over the Green's functions can be calculated in the same way as in Eq.~\eqref{sumG}. Inserting it into the above expression yields
\begin{equation}
\sigma_{1z}^L=\frac{s\eta_z^2e^2 k_0\alpha }{(2\pi)^3\hbar}\int d\varphi'\int d\kappa'\kappa'
\frac{1}{q^2_{xy}+\alpha^2q^2_z}\,.
\end{equation}
The function in the integral diverges at $\bm{q}=(q_{xy}, q_z)=0$. Transferring the
integral over $\kappa'$ to that over $\bm{q}$, we obtain
\begin{equation}
\begin{split}
\sigma_{1z}^L
&=\frac{s\eta_z^2e^2 k_0\alpha^2 }{(2\pi)^3\hbar}\int dq_{xy} dq_z
\frac{1}{q^2_{xy}+\alpha^2q^2_z}\\
&\underset{(q_z'=\alpha q_z)}{=}\frac{s\eta_z^2e^2 k_0\alpha}{(2\pi)^3\hbar}\int dq_{xy} dq_z'
\frac{1}{q^2_{xy}+q'^2_z}\\
&=\frac{s\eta_z^2e^2 k_0\alpha}{(2\pi)^2 h}\int^{1/\ell_e}_{1/\ell_\phi}
\frac{q'dq'}{{q'}^2},
\end{split}
\end{equation}
where ${q'}^2=q_{xy}^2+{q'_z}^2$ and the same cutoff of the integral as that in the SR limit is adopted.
Finally, we obtain
\begin{equation}
\sigma_{1z}^L=\frac{s\eta_z^2 \mathcal{K}\alpha e^2}{(2\pi)^2 h}\ln(\ell_\phi/\ell_e),
\end{equation}
where $\mathcal{K}=2\pi k_0$ is the circumference of the nodal loop.

\subsubsection{Dressed Hikami box in the LR limit}

Similarly to the SR limit, there are two dressed Hikami boxes contributing to the conductivity correction in the LR limit. The contribution by the retarded dressed Hikami box in the LR limit is
\begin{equation}
\begin{split}
\sigma_{2z}^L
&=s\eta_z^2\frac{e^2\hbar}{2\pi V^3}\sum_{\bm{k},\bm{k}',\bm{k}_1}v^z_{\delta\bm{k}}v^z_{\bm{q}-\delta\bm{k}_1}
G^A_{\delta\bm{k}}G^A_{\bm{q}-\delta\bm{k}_1}C_{\delta\bm{k}_1,\bm{q}-\delta\bm{k}}
G^R_{\delta\bm{k}}G^R_{\delta\bm{k}_1}
C^0_{\delta\bm{k},\delta\bm{k}_1}G^R_{\bm{q}-\delta\bm{k}}G^R_{\bm{q}-\delta\bm{k}_1}\\
&=s\eta_z^2\frac{e^2\hbar}{(2\pi)^4}\int d\theta'\int d\varphi'\int d\kappa'\kappa'\frac{k_0}{\alpha}\int\frac{d\theta}{2\pi}\int\frac{d\varphi}{2\pi}\int \rho(\varepsilon) d\varepsilon\int\frac{d\theta_1}{2\pi}\int\frac{d\varphi_1}{2\pi}\int \rho(\varepsilon_1) d\varepsilon \\
&\times v^z_{\delta\bm{k}}v^z_{\bm{q}-\delta\bm{k}_1}
G^A_{\delta\bm{k}}G^A_{\bm{q}-\delta\bm{k}_1}C_{\delta\bm{k}_1,\bm{q}-\delta\bm{k}}
G^R_{\delta\bm{k}}G^R_{\delta\bm{k}_1}
C^0_{\delta\bm{k},\delta\bm{k}_1}G^R_{\bm{q}-\delta\bm{k}}G^R_{\bm{q}-\delta\bm{k}_1}
\end{split}
\end{equation}
where $\bm{q}=\delta\bm{k}_1+\delta\bm{k}'$. The bare
Cooperon $C^0_{\delta\bm{k},\delta\bm{k}_1}$ restricts
$\bm{k}$ and $\bm{k}_1$ to the phase space $|\theta-\theta_1|<\Delta$.
As $\Delta\rightarrow0$, it means that
$\theta\simeq\theta_1$. The full Cooperon $C_{\delta\bm{k}_1,
\bm{q}-\delta\bm{k}}$ also restricts the
two momentums $\delta\bm{k}_1$ and $\delta\bm{k}'-\delta\bm{k}+\delta\bm{k}_1$
to the same plane. Since we already
have $\theta\simeq\theta_1$, we now further maintain
$\theta'\simeq\theta_1\simeq\theta$. As a result,
we have $\theta_q\simeq\theta$, which removes the factor $\cos^2(\theta_Q-\theta)$ in
the denominator of the full Cooperon in Eq. \eqref{coop2}.
The full Cooperon therefore reduces to $C_{\delta\bm{k}_1,
\bm{q}-\delta\bm{k}}\simeq\frac{f_\Delta(\theta-\theta')}{2}\frac{\gamma
e^{-i(\varphi_1-\varphi-\pi)}}{v_0^2q_{xy}^2\tau_e^2
+\alpha^2v_0^2q_z^2\tau_e^2}$.
Inserting the bare and full Cooperons and using the property that the full Cooperon
diverges as $\bm{q}\rightarrow0$ we have
\begin{equation}
\begin{split}
\sigma_{2z}^{L}
&=-s\eta_z^2\frac{e^2\hbar}{(2\pi)^2}\frac{k_0}{\alpha}\frac{(\alpha v_0 \gamma)^2}{8}\int \frac{d\theta'}{2\pi}\int \frac{d\varphi'}{2\pi}\int d\kappa'\kappa'\int\frac{d\theta}{2\pi}\int\frac{d\varphi}{2\pi}\int \rho(\varepsilon) d\varepsilon\int\frac{d\theta_1}{2\pi}\int\frac{d\varphi_1}{2\pi}\int \rho(\varepsilon_1) d\varepsilon_1 \sin\varphi\sin\varphi_1\\
&\times G^A_{\delta\bm{k}}G^A_{-\delta\bm{k}_1}G^R_{\delta\bm{k}}G^R_{\delta\bm{k}_1}
G^R_{-\delta\bm{k}}G^R_{-\delta\bm{k}_1}
\frac{f_\Delta(\theta_1-\theta)f_\Delta(\theta-\theta')\big[1+2e^{-i(\varphi-\varphi_1)}+e^{-2i
(\varphi-\varphi_1)}\big]
e^{-i(\varphi_1-\varphi-\pi)}}{v_0^2q_{xy}^2\tau_e^2
+\alpha^2v_0^2q_z^2\tau_e^2}\\
&=-s\eta_z^2\frac{e^2\hbar}{(2\pi)^2}\frac{k_0}{\alpha}\frac{(\alpha v_0 \gamma)^2}{8}\frac{\Delta^2}{\pi^2}\Big[\int\frac{d\varphi}{2\pi} \int\frac{d\varphi_1}{2\pi} \sin\varphi\sin\varphi_1\big[1+2e^{-i(\varphi-\varphi_1)}+e^{-2i
(\varphi-\varphi_1)}\big]
e^{-i(\varphi_1-\varphi-\pi)}\Big]_{=-1/2}\\
&\times\int \rho(\varepsilon) d\varepsilon\int \rho(\varepsilon_1) d\varepsilon_1
G^A_{\delta\bm{k}}G^A_{-\delta\bm{k}_1}G^R_{\delta\bm{k}}G^R_{\delta\bm{k}_1}
G^R_{-\delta\bm{k}}G^R_{-\delta\bm{k}_1}
\times\int\frac{d\varphi'}{2\pi}\int d\kappa'\kappa'\frac{1}{v_0^2q_{xy}^2\tau_e^2
+\alpha^2v_0^2q_z^2\tau_e^2}\\
&=s\eta_z^2\frac{e^2\hbar}{(2\pi)^3}\frac{k_0}{\alpha}\frac{(\alpha v_0 \gamma)^2}{16}\frac{\Delta^2}{\pi^2}\int \rho(\varepsilon) d\varepsilon G^A_{\delta\bm{k}}G^R_{\delta\bm{k}}G^R_{-\delta\bm{k}}\int \rho(\varepsilon_1) d\varepsilon_1
G^A_{-\delta\bm{k}_1}G^R_{\delta\bm{k}_1}G^R_{-\delta\bm{k}_1}\\
&\times\int d\varphi'\int d\kappa'\kappa'\frac{1}{v_0^2q_{xy}^2\tau_e^2
+\alpha^2v_0^2q_z^2\tau_e^2}\\
&=s\eta_z^2\frac{e^2\hbar}{(2\pi)^3}\frac{k_0}{\alpha}\frac{(\alpha v_0 \gamma)^2}{16}\frac{\Delta^2}{\pi^2}\Big(\int \rho(\varepsilon) d\varepsilon G^A_{\delta\bm{k}}G^R_{\delta\bm{k}}G^R_{-\delta\bm{k}}\Big)^2\int d\varphi'\int d\kappa'\kappa'\frac{1}{v_0^2q_{xy}^2\tau_e^2
+v_0^2q_z^2\tau_e^2}\\
&=-\frac{s\eta_z^2 e^2 k_0\alpha^2}{4(2\pi)^3\hbar}\int dq_{xy} dq_z\frac{1}{q_{xy}^2
+\alpha^2q_z^2}
=-\frac{1}{4}\sigma^L_{1z}.
\end{split}
\end{equation}
As a result, the correction due to each of the dressed Hikami boxes is also $-1/4$ of the bare one, similarly to the SR limit. The final quantum correction to the conductivity in the LR limit is
\begin{equation}
\sigma_z^{L}=\frac{1}{2}\sigma_{1z}^{L}=\frac{s\eta_z^2 \mathcal{K}\alpha e^2}{2(2\pi)^2 h}\ln(\ell_\phi/\ell_e),
\end{equation}
which is Eq. (7b) in the main text.

\subsection{Magnetoconductivity in the SR limit}
\begin{figure}[t!]

 \centering
                \includegraphics[width=0.4\columnwidth]{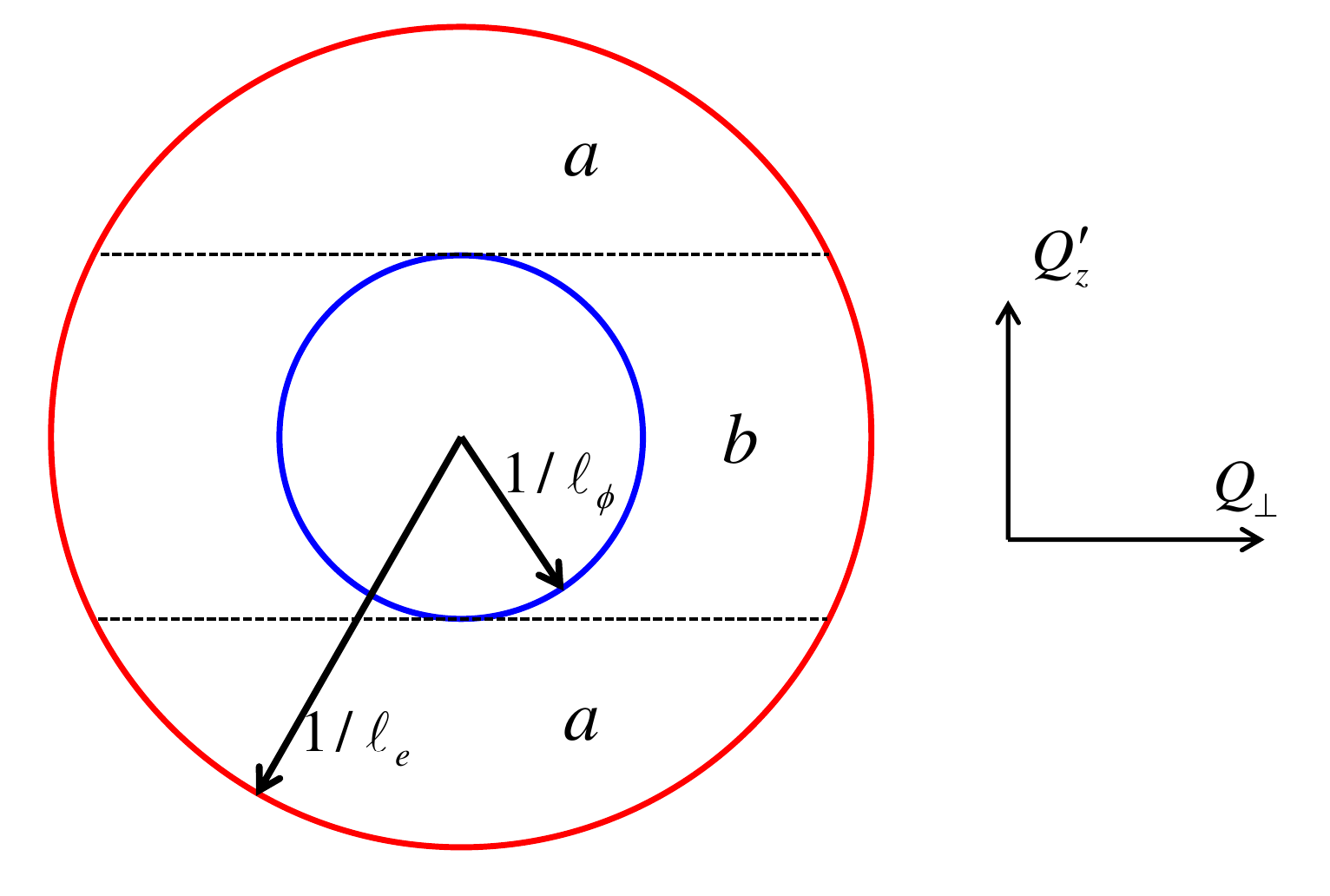}
\caption{The cutoff of the integral over $Q$ is the two cirlces with the radius being $1/\ell_e$ and $1/\ell_\phi$.
Correspondingly, the cutoff for the integral over $Q_\perp$ can be read from it and the cutoff depends on $Q_z$.
}
\label{cutoff}
\end{figure}

The full correction to the conductivity in the SR limit in the absence of magnetic field ($B=0$) [cf.~Eq.~\eqref{condwn}] can be rewritten as
\begin{equation}\label{b=0}
\begin{split}
\sigma_z^S(0)
&=-\frac{2s\eta_z^2\alpha^2 e^2}{h}\frac{1}{V}\sum_{\bm{Q}}\frac{1}{Q_\perp^2+4\alpha^2Q_z^2}\\
&=-\frac{2s\eta_z^2\alpha^2 e^2}{h}\frac{1}{(2\pi)^3}\int dQ_z\int 2\pi Q_\perp dQ_\perp\frac{1}{Q_\perp^2+4\alpha^2Q_z^2}\\
&=-\frac{s\eta_z^2\alpha^2 e^2}{(2\pi)^2 h}\int dQ_z\int\frac{dQ_\perp^2}{Q_\perp^2+4\alpha^2Q_z^2}\\
&=-\frac{s\eta_z^2\alpha e^2}{2(2\pi)^2 h}\int dQ'_z\int\frac{dQ_\perp^2}{Q_\perp^2+{Q'_z}^2}
\end{split}\,,
\end{equation}
where $Q_\perp=Q_{xy}$ is the momentum component perpendicular to the magnetic field in the $z$-direction.
The magnetic field introduces dephasing to the quantum interference, thus
suppressing the quantum correction to the conductivity and leads to
finite magnetoconductivity. This can be explained by Landau-level quantization of the quantum states,
imposing the quantization condition to $Q_\perp$
perpendicular to $B$ as
\begin{equation}
Q_\perp^2\rightarrow Q_n^2=(n+\frac{1}{2})\frac{4eB}{\hbar}=(n+\frac{1}{2})\frac{1}{\ell_{B}^2}\,.
\end{equation}
In the calculation,
we adopted the cutoff of the 3D momentum
$Q'=\sqrt{Q_\perp^2+{Q'_z}^2}$ in the integral in terms of $\ell_e$ and $\ell_\phi$.
In order to recover the result for $B=0$, here we also use the same
cutoff. As a result, the cutoff of $Q_\perp$
is a function of $Q'_z$.
When $Q'_z$ lies in the region a and b in Fig. \ref{cutoff}, $Q_\perp$ has different kinds of cutoff, which is
summarized below
\begin{equation}
\begin{split}
Q'^2&=Q_\perp^2+{Q'_z}^2\\
Q'^2_{\text{max}}&=1/\ell_e^2,\ \ \ Q'^2_{\text{min}}=1/\ell^2_\phi\\
(a)\ \ \  |Q'_z|\in(1/\ell_\phi,1/\ell_e):&\ \ \ Q_\perp^2\in(0,1/\ell_e^2-{Q'_z}^2)\\
n_{\text{max}}&=\ell_{B}^2/\ell_e^2-{Q'_z}^2\ell^2_{B}\\
n_{\text{min}}&=0\\
(b)\ \ \  |Q'_z|\in(0,1/\ell_\phi):&\ \ \ Q_\perp^2\in(1/\ell_\phi^2-{Q'_z}^2,1/\ell_e^2-{Q'_z}^2)\\
n_{\text{max}}&=\ell_{B}^2/\ell_e^2-{Q'_z}^2\ell^2_{B}\\
n_{\text{min}}&=\ell_{B}^2/\ell_\phi^2-{Q'_z}^2\ell^2_{B}\\
\end{split}\,.
\end{equation}
We use the Dirac delta function to rewrite Eq. \eqref{b=0} with finite $B$ as
\begin{equation}
\begin{split}
\sigma_z^S(B)&=-\frac{s\eta_z^2\alpha e^2}{2(2\pi)^2 h}\Big\{ 2\int_0^{1/\ell_\phi} dQ'_z\sum_n\int_{1/\ell_{\phi}^2-{Q'_z}^2}^{1/\ell_e^2-{Q'_z}^2}
\frac{dQ_\perp^2}{Q_\perp^2+{Q'_z}^2}\delta\big[(n+\frac{1}{2})-\ell_{B}^2Q_\perp^2\big]\\
&+2\int_{1/\ell_\phi}^{1/\ell_e} dQ'_z\sum_n\int_{0}^{1/\ell_e^2-{Q'_z}^2}
\frac{dQ_\perp^2}{Q_\perp^2+{Q'_z}^2}\delta\big[(n+\frac{1}{2})-\ell_{B}^2Q_\perp^2\big]\Big\}\\
&=-\frac{s\eta_z^2\alpha e^2}{2(2\pi)^2 h}\Big\{
2\int_0^{1/\ell_\phi}dQ'_z\sum_{n=\ell_{B}^2/\ell_\phi^2-\ell_{B}^2{Q'_z}^2}^{\ell_{B}^2/\ell_e^2-\ell_{B}^2{Q'_z}^2}
\frac{1}{n+\frac{1}{2}+\ell_{B}^2{Q'_z}^2}\\
&+2\int_{1/\ell_\phi}^{1/\ell_e}d{Q}'_z\sum_{n=0}^{\ell_{B}^2/\ell_e^2-\ell_{B}^2{Q'_z}^2}
\frac{1}{n+\frac{1}{2}+\ell_{B}^2{Q'_z}^2}\Big\}.
\end{split}
\end{equation}
Using the property of the digamma function $\Psi(x)$:
\begin{equation}
\Psi(x+N)-\Psi(x)=\sum_{n=0}^{N-1}\frac{1}{x+n},
\end{equation}
the magnetoconductivity reduces to
\begin{equation}
\begin{split}
\sigma_z^S(B)&=-\frac{s\eta_z^2\alpha e^2}{(2\pi)^2 h}\Big[
\Psi(\ell^2_{B}/\ell_e^2+\frac{1}{2})/\ell_e
-\Psi(\ell^2_{B}/\ell_\phi^2+\frac{1}{2})/\ell_\phi
-\int_{1/\ell_\phi}^{1/\ell_e}dx\Psi(\ell_{B}^2x^2+\frac{1}{2})\Big],
\end{split}
\end{equation}
which is Eq. (8a) in the main text.
In the zero field limit $B\rightarrow0, \ell_{B}\rightarrow\infty$, using $\Psi(x+\frac{1}{2})\simeq\ln x$ for $x\rightarrow\infty$, the above expression reduces to
\begin{equation}
\sigma_z^S(B=0)=-\frac{s\eta_z^2\alpha e^2}{2\pi^2 h}(\frac{1}{\ell_e}-\frac{1}{\ell_\phi}),
\end{equation}
which is Eq. (7a) in the main text. The magnetoconductivity is defined as
$\delta\sigma^S(B)\equiv\sigma_z^S(B)-\sigma_z^S(0)$, which can be probed in experiments.

\subsection{Magnetoconductivity in the LR limit}
Note that the interference loops in the LR limit
are confined within various planes parallel to the $z$-axis,
so that electrons cannot feel any flux when the external
magnetic flux is in the $z$-direction. Therefore,
here, we impose a magnetic field in the $x$-direction.
In this case, electrons in different interference loops feel a different
enclosed flux. For the plane denoted by $\theta$,
the component of the magnetic field perpendicular to the
plane is $B(\theta)=B\sin\theta$. In the $\theta$-plane,
the Hamiltonian for the 2D subsystem is
\begin{equation}
H_{2D}=\hbar v_0(\delta k_{xy}\sigma_x+\alpha\delta k_z\sigma_y),
\end{equation}
which contains an anisotropic factor $\alpha$.
In the presence of a magnetic field, the Hamiltonian in real space becomes (for simplicity, we set
the two directions of $\delta k_{xy}$ and $\delta k_z$ to the $x_1$ and the $y_1$ axes)
\begin{equation}
H_{2D}=\hbar v_0\big[(-i\partial_{x_1}-eBy_1/\hbar)\sigma_x+\alpha(-i\partial_{y_1})\sigma_y\big].
\end{equation}
When $\alpha=1$, the quantization condition for the Cooperon is
$q_{x_1}^2+q_{y_1}^2=(n+\frac{1}{2})\frac{4e|B(\theta)|}{\hbar}$.
For the anisotropic case, by performing the substitution $x_1\rightarrow \tilde{x}/\sqrt{\alpha}, y_1\rightarrow \sqrt{\alpha}\tilde{y}$, the Hamiltonian reduces to
\begin{equation}
\tilde{H}_{2D}=\sqrt{\alpha}\hbar v_0\big[(-i\partial_{\tilde{x}}-eB\tilde{y}/\hbar)\sigma_x+(-i\partial_{\tilde{y}})\sigma_y\big].
\end{equation}
For this isotropic Dirac Hamiltonian, the quantization condition is
\begin{equation}
\tilde{q}_x^2+\tilde{q}_y^2=(n+\frac{1}{2})\frac{4e|B(\theta)|}{\hbar}.
\end{equation}
Performing the inverse substitution $\tilde{q}_x\rightarrow q_{x_1}/\sqrt{\alpha}, \tilde{q}_y\rightarrow \sqrt{\alpha}q_{y_1}$
then we obtain the quantization condition for the momentum
\begin{equation}
q_{x_1}^2+\alpha^2 q_{y_1}^2=(n+\frac{1}{2})\frac{4\alpha e|B(\theta)|}{\hbar}.
\end{equation}
We can rewrite the expression in the initial coordinates system
\begin{equation}
q'^2=q_{xy}^2+\alpha^2 q_z^2=(n+\frac{1}{2})\frac{4\alpha e|B(\theta)|}{\hbar}=(n+\frac{1}{2})\frac{\alpha|\sin\theta|}{l_{B}^2}.
\end{equation}
The upper and lower cutoff of the series is
\begin{equation}
\begin{split}
q'^2_{\text{max}}=1/\ell_e^2, \ \ q'^2_{\text{min}}=1/\ell_\phi^2\\
n_{\text{max}}=\ell_{B}^2/(\ell_e^2\alpha|\sin\theta|)\\
n_{\text{min}}=\ell_{B}^2/(\ell_\phi^2\alpha|\sin\theta|)
\end{split}\,.
\end{equation}

Now, the magnetic field has different contributions for different 2D diffusion planes.
Starting with the full correction to the conductivity which is $\sigma_{1z}^L/2$ [cf.~Eq.~\eqref{cond2}] and imposing the quantization condition for $q'$
\begin{equation}
\begin{split}
\sigma_z^{L}(B)
&=s\eta_z^2\frac{e^2\hbar}{8\pi V^2}\frac{k_0\Delta}{\alpha}\frac{n_iu_0^2}{\hbar^2v_0^2\tau_e^2}\sum_{\bm{k}} {v^z_{\bm{k}}}^2 {G^R_{\bm{k}}}^2{G^A_{\bm{k}}}^2
\times \frac{V}{(2\pi)^2}\sum_n\int_{1/\ell_\phi^2}^{1/\ell_e^2} \frac{d(q'^2)}{q'^2}\delta\big[(n+\frac{1}{2})-q'^2\ell_{B}^2/(\alpha|\sin\theta|)\big]\\
&=s\eta_z^2\frac{e^2\hbar}{8\pi V^2}\frac{k_0\Delta}{\alpha}\frac{n_iu_0^2}{\hbar^2v_0^2\tau_e^2}\sum_{\bm{k}} {v^z_{\bm{k}}}^2 {G^R_{\bm{k}}}^2{G^A_{\bm{k}}}^2
\times \frac{V}{(2\pi)^2}\sum_{n=\ell_{B}^2/(\ell_\phi^2\alpha|\sin\theta|)}^{\ell_{B}^2/(\ell_e^2\alpha|\sin\theta|)} \frac{1}{n+\frac{1}{2}}\\
&=s\eta_z^2\frac{e^2\hbar}{8\pi V^2}\frac{k_0\Delta}{\alpha}\frac{n_iu_0^2}{\hbar^2v_0^2\tau_e^2}\sum_{\bm{k}} {v^z_{\bm{k}}}^2 {G^R_{\bm{k}}}^2{G^A_{\bm{k}}}^2
\times \frac{V}{(2\pi)^2}\Big[\Psi(\frac{\ell_{B}^2}{\ell_e^2\alpha|\sin\theta|}+\frac{1}{2})
-\Psi(\frac{\ell_{B}^2}{\ell_\phi^2\alpha|\sin\theta|}+\frac{1}{2})\Big]
\end{split}\,,
\end{equation}
then we obtain
\begin{equation}
\sigma_z^L(B)=\frac{s\eta_z^2 \mathcal{K}\alpha e^2}{16\pi^2h}
\int^{2\pi}_0\frac{d\theta}{2\pi}
\Big[\Psi(\frac{\ell_{B}^2}{\ell_e^2\alpha|\sin\theta|}+\frac{1}{2})
-\Psi(\frac{\ell_{B}^2}{\ell_\phi^2\alpha|\sin\theta|}+\frac{1}{2})\Big],
\end{equation}
which is Eq. (8b) in the main text.
As $B\rightarrow0$, $\ell_{B}\rightarrow\infty$, the above result
reduces to
\begin{equation}
\begin{split}
\sigma_z^L(B=0)=\frac{s\eta_z^2 \mathcal{K}\alpha e^2}{2(2\pi)^2h}
\ln(\ell_\phi/\ell_e)\,,
\end{split}
\end{equation}
which recovers the zero-field result Eq. (7b) in the main text.
The magnetoconductivity is defined in the same way as in the SR limit as
$\delta\sigma^L(B)\equiv\sigma_z^L(B)-\sigma_z^L(0)$.

\subsection{Effects of finite temperature,
spin-orbit coupling, electron-electron interaction, tilting of the nodal line, and Berry curvature}

Here we discuss possible extensions of our theory that include other effects on the transport property, including finite temperature, spin-orbit coupling, electron-electron interaction, tilted nodal line, and Berry curvature.\\
\\
(1) \emph{Temperature dependence of the conductivity}.
At sufficiently low temperatures in the
measurement of the WL effect, the
temperature dependence of the quantum correction
to the conductivity mainly arises from the
variation of the phase coherence length $\ell_\phi$.
Electron-electron and electron-phonon interaction
are two dominant decoherence mechanisms.
Both scenarios result in a
temperature dependence of the coherence length as $\ell_\phi=CT^{-p/2}$
with different coefficients $C$ \cite{Akkermans07}.
In 3D, $p=3/2\ (p=3)$ for the electron-electron (electron-phonon)
interaction-induced decoherence. At low temperature, the electron-electron interaction
dominates the decoherence effect \cite{Datta97book}, so that by
inserting $\ell_\phi=CT^{-3/4}$ into Eqs. (7a) and (7b) in the main text,
we obtain the temperature dependent quantum
correction to the conductivity
\begin{subequations}\label{cond}
\begin{align}
\sigma_z^{S}&=-\frac{s\eta_z^2\alpha e^2}{2\pi^2h}(\frac{1}{\ell_e}-\frac{T^{3/4}}{C}),\label{cond1}\\
\sigma_z^{L}&=\frac{s\eta_z^2 \mathcal{K}\alpha e^2 }{2(2\pi)^2 h}\ln[C/(\ell_eT^{3/4})], \label{cond2}
\end{align}
\end{subequations}
\\
(2) \emph{Spin-orbit coupling}. The model in our work is a two-band
model with spin degeneracy, which is the case for most nodal
line semimetals. In these materials, spin-orbit coupling is
a small perturbation and will not modify the band structure, i.e.,
the real spin and pseudo-spin are decoupled. We also
assume that the overall effect due to spin-orbit coupling
results solely in spin relaxation, and will not change the interference
trajectory determined by the disorder in both limiting cases.
In this case, the effect of spin-orbit coupling on our predicted WL/WAL
can be analyzed in a similar way to that employed in
conventional metals: the Cooperon contains spin-resolved
structure factors, which can be diagonalized into one singlet
and three triplet channels \cite{Akkermans07}. Without spin-orbit coupling,
the singlet channel contributes $-1/2$ of the quantum correction
to the conductivity while each of the triplet channels contributes a factor
$1/2$ such that all together the triplet channels contribute $3/2$. Therefore, the
overall factor of the correction is 1 and is equivalent to
summing over each spin channel individually (as reported in our main text).
The impact of spin-orbit coupling is to
suppress the triplet channels while the singlet channel
contribution remains unaffected, because spin-orbit coupling
does not break time-reversal symmetry. Taking the spin-orbit
coupling sufficiently strong, such that the spin-relaxation time
is much smaller than the phase coherence time, $\tau_{so}\ll\tau_\phi$, the triplet
channels will be completely suppressed and will not contribute
to the quantum correction. As a result, the overall factor of the
correction to the conductivity becomes $-1/2$, which results
in a sign change as well as a reduction of the magnitude (with
respect to our reported result). Applying
this contribution to our results, we expect that as
the strength of the spin-orbit coupling increases, the system will
undergo a crossover from 3D WL to 3D WAL in the SR limit and from 2D
WAL to 2D WL in the LR
limit. Such crossover behaviors can be revealed by the magnetoconductivity measurements.\\
\\
(3) \emph{Electron-electron interaction}. The consequences
of electron-electron interaction
can be classified into two categories \cite{Akkermans07}:
(i) decoherence and (ii) Altshuler-Aronov effect \cite{Altshuler85}.
In the following, we discuss both effects:\\
\\
(i) The interaction
between electrons is an inelastic process that destroys
their phase coherence after a characteristic time, i.e., it
leads to decoherence. Each
electron moves through electronic density fluctuations
induced by other electrons, in similitude to a fluctuating
electromagnetic field that dephases the coherent trajectories
of moving electrons. As a result, the phase coherence
length $\ell_\phi$ becomes temperature-dependent.
This effect is included in the quantum
correction to the conductivity [Eqs. (7a), (7b) in the main text],
and can be measured in magnetoconductivity signals
[Eqs. (8a), (8b) in the main text].\\
\\
(ii) The second consequence of interaction is
the localization-like behavior due to the interplay
between electron-electron interaction and disorder
scattering (known as the Altshuler-Aronov effect \cite{Altshuler85}). The
diffusive motion of electrons enhances the interaction
effect and causes a change of the density of states
as well as a change to the conductivity around the Fermi level \cite{Akkermans07}.
For conventional metals and Weyl semimetals, the
change in conductivity is of the same order of
magnitude as the WL/WAL correction, but its origin and
nature are quite different. In particular, the interaction-induced
conductivity correction (Altshuler-Aronov effect)
weakly depends on the magnetic field, compared
with the WL/WAL corrections. Based on its
different origin, the usual way to extract the former
correction is to measure the
temperature dependence of the conductivity alongside
quenching the latter quantum correction by applying a
magnetic field.
An interesting point for the interaction correction
to the conductivity is that its temperature dependence
also shows the dimensionality effect, similar to our
results for the WL/WAL effect. Therefore, one can
expect different temperature dependence of the
interaction correction to the conductivity in
different types of impurity potentials.\\
\\
(4) \emph{Tilting of the nodal line}. The tilting
of the nodal line may lead to two main effects, the
tilting of the band dispersion in the transverse
directions of the nodal line \cite{Li17prb} and
the finite dispersion along the nodal line \cite{Ahn17prl,Huang18prb}.
The effects of such tilts on the quantum correction to the conductivity
can be analyzed through symmetry arguments. When the
tilting breaks $T_1$ symmetry, it will suppress the 3D
WL effect; similarly, when the tilting
breaks $T_2$ symmetry, it will suppress the 2D WAL effect.
The $T_1$ or $T_2$ symmetry-breaking
can be regarded as dephasing effects, that suppress the
quantum correction to the conductivity. To be specific,
when only the former tilting exists, the $T_2$ symmetry
is broken while the $T_1$ symmetry remains. As a result, the
2D WAL is suppressed while the
3D WL survives. For the second kind of
tilting, finite dispersion is introduced along the nodal
line, which may or may not break $T_1$ symmetry,
i.e., depending on the details on the tilting it can suppress the 3D WL.
When both $T_1$ and $T_2$ symmetries are broken,
both the WL and WAL
effects are suppressed. Such a symmetry analysis
should be combined with the consideration of the
dominant quantum interference processes in both limits of the impurity potential
in order to generalize our results to these perturbations.\\
\\
(5) \emph{Quantum anomaly and Berry curvature effect}.
The quantum anomaly in the nodal line semimetal is a
parity anomaly \cite{Rui18prb,Burkov18prb}, which is different
from the chiral anomaly in Weyl/Dirac semimetals. The chiral anomaly
in Weyl/Dirac semimetals manifests
itself in negative magnetoresistivity. The parity anomaly
in the nodal line semimetal is predicted to induce anomalous
transverse current, but there is no magnetoresistivity
expected \cite{Rui18prb}.
Aside from the chiral anomaly in Weyl/Dirac semimetals,
classical negative magnetoresistivity may also be induced
by a Berry curvature effect \cite{Son13prb,Dai17prl}.
Such an effect has yet to be explored in nodal line semimetals.
Different from the WL/WAL effect, such a classical
magnetoresistivity possesses a $B^2$-dependence of the
magnetic field, and will be overwhelmed by the 3D WL ($\sim\sqrt{B}$) or 2D
WAL ($\sim\ln B$) effects in the weak-field limit. Moreover,
such a classical effect is not sensitive to temperature,
which is another distinction compared with the WL and
WAL effects. Based on the observation above, we
conclude that: (i) At low temperatures, the
WL/WAL effects and classical Berry curvature effect
dominate the magnetoconductivity in the weak and
strong magnetic field limits, respectively. (ii)
At high temperature, the WL/WAL correction is
quenched and the Berry curvature effect dominate
the magnetoconductivity. Therefore, the WL/WAL
and Berry curvature contribution to the magnetoconductivity
can be distinguished by both their magnetic field and temperature dependence.


\begin{thebibliography}{68}%
\makeatletter
\providecommand \@ifxundefined [1]{%
 \@ifx{#1\undefined}
}%
\providecommand \@ifnum [1]{%
 \ifnum #1\expandafter \@firstoftwo
 \else \expandafter \@secondoftwo
 \fi
}%
\providecommand \@ifx [1]{%
 \ifx #1\expandafter \@firstoftwo
 \else \expandafter \@secondoftwo
 \fi
}%
\providecommand \natexlab [1]{#1}%
\providecommand \enquote  [1]{``#1''}%
\providecommand \bibnamefont  [1]{#1}%
\providecommand \bibfnamefont [1]{#1}%
\providecommand \citenamefont [1]{#1}%
\providecommand \href@noop [0]{\@secondoftwo}%
\providecommand \href [0]{\begingroup \@sanitize@url \@href}%
\providecommand \@href[1]{\@@startlink{#1}\@@href}%
\providecommand \@@href[1]{\endgroup#1\@@endlink}%
\providecommand \@sanitize@url [0]{\catcode `\\12\catcode `\$12\catcode
  `\&12\catcode `\#12\catcode `\^12\catcode `\_12\catcode `\%12\relax}%
\providecommand \@@startlink[1]{}%
\providecommand \@@endlink[0]{}%
\providecommand \url  [0]{\begingroup\@sanitize@url \@url }%
\providecommand \@url [1]{\endgroup\@href {#1}{\urlprefix }}%
\providecommand \urlprefix  [0]{URL }%
\providecommand \Eprint [0]{\href }%
\providecommand \doibase [0]{http://dx.doi.org/}%
\providecommand \selectlanguage [0]{\@gobble}%
\providecommand \bibinfo  [0]{\@secondoftwo}%
\providecommand \bibfield  [0]{\@secondoftwo}%
\providecommand \translation [1]{[#1]}%
\providecommand \BibitemOpen [0]{}%
\providecommand \bibitemStop [0]{}%
\providecommand \bibitemNoStop [0]{.\EOS\space}%
\providecommand \EOS [0]{\spacefactor3000\relax}%
\providecommand \BibitemShut  [1]{\csname bibitem#1\endcsname}%
\let\auto@bib@innerbib\@empty
\bibitem [{\citenamefont {Anderson}(1958)}]{Anderson58pr}%
  \BibitemOpen
  \bibfield  {author} {\bibinfo {author} {\bibfnamefont {P.~W.}\ \bibnamefont
  {Anderson}},\ }\href {\doibase 10.1103/PhysRev.109.1492} {\bibfield
  {journal} {\bibinfo  {journal} {Phys. Rev.}\ }\textbf {\bibinfo {volume}
  {109}},\ \bibinfo {pages} {1492} (\bibinfo {year} {1958})}\BibitemShut
  {NoStop}%
\bibitem [{\citenamefont {Lee}\ and\ \citenamefont
  {Ramakrishnan}(1985)}]{Lee85rmp}%
  \BibitemOpen
  \bibfield  {author} {\bibinfo {author} {\bibfnamefont {P.~A.}\ \bibnamefont
  {Lee}}\ and\ \bibinfo {author} {\bibfnamefont {T.~V.}\ \bibnamefont
  {Ramakrishnan}},\ }\href {\doibase 10.1103/RevModPhys.57.287} {\bibfield
  {journal} {\bibinfo  {journal} {Rev. Mod. Phys.}\ }\textbf {\bibinfo {volume}
  {57}},\ \bibinfo {pages} {287} (\bibinfo {year} {1985})}\BibitemShut
  {NoStop}%
\bibitem [{\citenamefont {Akkermans}\ and\ \citenamefont
  {Montambaux}(2007)}]{Akkermans07}%
  \BibitemOpen
  \bibfield  {author} {\bibinfo {author} {\bibfnamefont {E.}~\bibnamefont
  {Akkermans}}\ and\ \bibinfo {author} {\bibfnamefont {G.}~\bibnamefont
  {Montambaux}},\ }\href@noop {} {\emph {\bibinfo {title} {Mesoscopic physics
  of electrons and photons}}}\ (\bibinfo  {publisher} {Cambridge university
  press},\ \bibinfo {year} {2007})\BibitemShut {NoStop}%
\bibitem [{\citenamefont {Dyson}(1962)}]{Dyson62jmp}%
  \BibitemOpen
  \bibfield  {author} {\bibinfo {author} {\bibfnamefont {F.~J.}\ \bibnamefont
  {Dyson}},\ }\href@noop {} {\bibfield  {journal} {\bibinfo  {journal} {Journal
  of Mathematical Physics}\ }\textbf {\bibinfo {volume} {3}},\ \bibinfo {pages}
  {140} (\bibinfo {year} {1962})}\BibitemShut {NoStop}%
\bibitem [{\citenamefont {Hikami}\ \emph {et~al.}(1980)\citenamefont {Hikami},
  \citenamefont {Larkin},\ and\ \citenamefont {Nagaoka}}]{Hikami80ptp}%
  \BibitemOpen
  \bibfield  {author} {\bibinfo {author} {\bibfnamefont {S.}~\bibnamefont
  {Hikami}}, \bibinfo {author} {\bibfnamefont {A.~I.}\ \bibnamefont {Larkin}},
  \ and\ \bibinfo {author} {\bibfnamefont {Y.}~\bibnamefont {Nagaoka}},\
  }\href@noop {} {\bibfield  {journal} {\bibinfo  {journal} {Progress of
  Theoretical Physics}\ }\textbf {\bibinfo {volume} {63}},\ \bibinfo {pages}
  {707} (\bibinfo {year} {1980})}\BibitemShut {NoStop}%
\bibitem [{\citenamefont {Hasan}\ and\ \citenamefont {Kane}(2010)}]{Kane10rmp}%
  \BibitemOpen
  \bibfield  {author} {\bibinfo {author} {\bibfnamefont {M.~Z.}\ \bibnamefont
  {Hasan}}\ and\ \bibinfo {author} {\bibfnamefont {C.~L.}\ \bibnamefont
  {Kane}},\ }\href {\doibase 10.1103/RevModPhys.82.3045} {\bibfield  {journal}
  {\bibinfo  {journal} {Rev. Mod. Phys.}\ }\textbf {\bibinfo {volume} {82}},\
  \bibinfo {pages} {3045} (\bibinfo {year} {2010})}\BibitemShut {NoStop}%
\bibitem [{\citenamefont {Qi}\ and\ \citenamefont {Zhang}(2011)}]{Qi11rmp}%
  \BibitemOpen
  \bibfield  {author} {\bibinfo {author} {\bibfnamefont {X.-L.}\ \bibnamefont
  {Qi}}\ and\ \bibinfo {author} {\bibfnamefont {S.-C.}\ \bibnamefont {Zhang}},\
  }\href {\doibase 10.1103/RevModPhys.83.1057} {\bibfield  {journal} {\bibinfo
  {journal} {Rev. Mod. Phys.}\ }\textbf {\bibinfo {volume} {83}},\ \bibinfo
  {pages} {1057} (\bibinfo {year} {2011})}\BibitemShut {NoStop}%
\bibitem [{\citenamefont {Ozawa}\ \emph {et~al.}(2018)\citenamefont {Ozawa},
  \citenamefont {Price}, \citenamefont {Amo}, \citenamefont {Goldman},
  \citenamefont {Hafezi}, \citenamefont {Lu}, \citenamefont {Rechtsman},
  \citenamefont {Schuster}, \citenamefont {Simon}, \citenamefont {Zilberberg}
  \emph {et~al.}}]{Ozawa18arxiv}%
  \BibitemOpen
  \bibfield  {author} {\bibinfo {author} {\bibfnamefont {T.}~\bibnamefont
  {Ozawa}}, \bibinfo {author} {\bibfnamefont {H.~M.}\ \bibnamefont {Price}},
  \bibinfo {author} {\bibfnamefont {A.}~\bibnamefont {Amo}}, \bibinfo {author}
  {\bibfnamefont {N.}~\bibnamefont {Goldman}}, \bibinfo {author} {\bibfnamefont
  {M.}~\bibnamefont {Hafezi}}, \bibinfo {author} {\bibfnamefont
  {L.}~\bibnamefont {Lu}}, \bibinfo {author} {\bibfnamefont {M.}~\bibnamefont
  {Rechtsman}}, \bibinfo {author} {\bibfnamefont {D.}~\bibnamefont {Schuster}},
  \bibinfo {author} {\bibfnamefont {J.}~\bibnamefont {Simon}}, \bibinfo
  {author} {\bibfnamefont {O.}~\bibnamefont {Zilberberg}},  \emph {et~al.},\
  }\href@noop {} {\bibfield  {journal} {\bibinfo  {journal} {arXiv preprint
  arXiv:1802.04173}\ } (\bibinfo {year} {2018})}\BibitemShut {NoStop}%
\bibitem [{\citenamefont {Suzuura}\ and\ \citenamefont
  {Ando}(2002)}]{Suzuura02prl}%
  \BibitemOpen
  \bibfield  {author} {\bibinfo {author} {\bibfnamefont {H.}~\bibnamefont
  {Suzuura}}\ and\ \bibinfo {author} {\bibfnamefont {T.}~\bibnamefont {Ando}},\
  }\href {\doibase 10.1103/PhysRevLett.89.266603} {\bibfield  {journal}
  {\bibinfo  {journal} {Phys. Rev. Lett.}\ }\textbf {\bibinfo {volume} {89}},\
  \bibinfo {pages} {266603} (\bibinfo {year} {2002})}\BibitemShut {NoStop}%
\bibitem [{\citenamefont {McCann}\ \emph {et~al.}(2006)\citenamefont {McCann},
  \citenamefont {Kechedzhi}, \citenamefont {Fal'ko}, \citenamefont {Suzuura},
  \citenamefont {Ando},\ and\ \citenamefont {Altshuler}}]{McCann06prl}%
  \BibitemOpen
  \bibfield  {author} {\bibinfo {author} {\bibfnamefont {E.}~\bibnamefont
  {McCann}}, \bibinfo {author} {\bibfnamefont {K.}~\bibnamefont {Kechedzhi}},
  \bibinfo {author} {\bibfnamefont {V.~I.}\ \bibnamefont {Fal'ko}}, \bibinfo
  {author} {\bibfnamefont {H.}~\bibnamefont {Suzuura}}, \bibinfo {author}
  {\bibfnamefont {T.}~\bibnamefont {Ando}}, \ and\ \bibinfo {author}
  {\bibfnamefont {B.~L.}\ \bibnamefont {Altshuler}},\ }\href {\doibase
  10.1103/PhysRevLett.97.146805} {\bibfield  {journal} {\bibinfo  {journal}
  {Phys. Rev. Lett.}\ }\textbf {\bibinfo {volume} {97}},\ \bibinfo {pages}
  {146805} (\bibinfo {year} {2006})}\BibitemShut {NoStop}%
\bibitem [{\citenamefont {Lu}\ \emph {et~al.}(2011)\citenamefont {Lu},
  \citenamefont {Shi},\ and\ \citenamefont {Shen}}]{Lu11prl}%
  \BibitemOpen
  \bibfield  {author} {\bibinfo {author} {\bibfnamefont {H.-Z.}\ \bibnamefont
  {Lu}}, \bibinfo {author} {\bibfnamefont {J.}~\bibnamefont {Shi}}, \ and\
  \bibinfo {author} {\bibfnamefont {S.-Q.}\ \bibnamefont {Shen}},\ }\href
  {\doibase 10.1103/PhysRevLett.107.076801} {\bibfield  {journal} {\bibinfo
  {journal} {Phys. Rev. Lett.}\ }\textbf {\bibinfo {volume} {107}},\ \bibinfo
  {pages} {076801} (\bibinfo {year} {2011})}\BibitemShut {NoStop}%
\bibitem [{\citenamefont {Garate}\ and\ \citenamefont
  {Glazman}(2012)}]{Garate12prb}%
  \BibitemOpen
  \bibfield  {author} {\bibinfo {author} {\bibfnamefont {I.}~\bibnamefont
  {Garate}}\ and\ \bibinfo {author} {\bibfnamefont {L.}~\bibnamefont
  {Glazman}},\ }\href {\doibase 10.1103/PhysRevB.86.035422} {\bibfield
  {journal} {\bibinfo  {journal} {Phys. Rev. B}\ }\textbf {\bibinfo {volume}
  {86}},\ \bibinfo {pages} {035422} (\bibinfo {year} {2012})}\BibitemShut
  {NoStop}%
\bibitem [{\citenamefont {Lu}\ and\ \citenamefont {Shen}(2014)}]{Lu14prl}%
  \BibitemOpen
  \bibfield  {author} {\bibinfo {author} {\bibfnamefont {H.-Z.}\ \bibnamefont
  {Lu}}\ and\ \bibinfo {author} {\bibfnamefont {S.-Q.}\ \bibnamefont {Shen}},\
  }\href {\doibase 10.1103/PhysRevLett.112.146601} {\bibfield  {journal}
  {\bibinfo  {journal} {Phys. Rev. Lett.}\ }\textbf {\bibinfo {volume} {112}},\
  \bibinfo {pages} {146601} (\bibinfo {year} {2014})}\BibitemShut {NoStop}%
\bibitem [{\citenamefont {Lu}\ and\ \citenamefont {Shen}(2015)}]{Lu15prb}%
  \BibitemOpen
  \bibfield  {author} {\bibinfo {author} {\bibfnamefont {H.-Z.}\ \bibnamefont
  {Lu}}\ and\ \bibinfo {author} {\bibfnamefont {S.-Q.}\ \bibnamefont {Shen}},\
  }\href {\doibase 10.1103/PhysRevB.92.035203} {\bibfield  {journal} {\bibinfo
  {journal} {Phys. Rev. B}\ }\textbf {\bibinfo {volume} {92}},\ \bibinfo
  {pages} {035203} (\bibinfo {year} {2015})}\BibitemShut {NoStop}%
\bibitem [{\citenamefont {Dai}\ \emph {et~al.}(2016)\citenamefont {Dai},
  \citenamefont {Lu}, \citenamefont {Shen},\ and\ \citenamefont
  {Yao}}]{Dai16prb}%
  \BibitemOpen
  \bibfield  {author} {\bibinfo {author} {\bibfnamefont {X.}~\bibnamefont
  {Dai}}, \bibinfo {author} {\bibfnamefont {H.-Z.}\ \bibnamefont {Lu}},
  \bibinfo {author} {\bibfnamefont {S.-Q.}\ \bibnamefont {Shen}}, \ and\
  \bibinfo {author} {\bibfnamefont {H.}~\bibnamefont {Yao}},\ }\href {\doibase
  10.1103/PhysRevB.93.161110} {\bibfield  {journal} {\bibinfo  {journal} {Phys.
  Rev. B}\ }\textbf {\bibinfo {volume} {93}},\ \bibinfo {pages} {161110}
  (\bibinfo {year} {2016})}\BibitemShut {NoStop}%
\bibitem [{\citenamefont {Ando}\ \emph {et~al.}(1998)\citenamefont {Ando},
  \citenamefont {Nakanishi},\ and\ \citenamefont {Saito}}]{Ando98jpsj}%
  \BibitemOpen
  \bibfield  {author} {\bibinfo {author} {\bibfnamefont {T.}~\bibnamefont
  {Ando}}, \bibinfo {author} {\bibfnamefont {T.}~\bibnamefont {Nakanishi}}, \
  and\ \bibinfo {author} {\bibfnamefont {R.}~\bibnamefont {Saito}},\
  }\href@noop {} {\bibfield  {journal} {\bibinfo  {journal} {Journal of the
  Physical Society of Japan}\ }\textbf {\bibinfo {volume} {67}},\ \bibinfo
  {pages} {2857} (\bibinfo {year} {1998})}\BibitemShut {NoStop}%
\bibitem [{\citenamefont {Burkov}\ \emph {et~al.}(2011)\citenamefont {Burkov},
  \citenamefont {Hook},\ and\ \citenamefont {Balents}}]{Burkov11prb}%
  \BibitemOpen
  \bibfield  {author} {\bibinfo {author} {\bibfnamefont {A.~A.}\ \bibnamefont
  {Burkov}}, \bibinfo {author} {\bibfnamefont {M.~D.}\ \bibnamefont {Hook}}, \
  and\ \bibinfo {author} {\bibfnamefont {L.}~\bibnamefont {Balents}},\ }\href
  {\doibase 10.1103/PhysRevB.84.235126} {\bibfield  {journal} {\bibinfo
  {journal} {Phys. Rev. B}\ }\textbf {\bibinfo {volume} {84}},\ \bibinfo
  {pages} {235126} (\bibinfo {year} {2011})}\BibitemShut {NoStop}%
\bibitem [{\citenamefont {Kim}\ \emph {et~al.}(2015)\citenamefont {Kim},
  \citenamefont {Wieder}, \citenamefont {Kane},\ and\ \citenamefont
  {Rappe}}]{Kim15prl}%
  \BibitemOpen
  \bibfield  {author} {\bibinfo {author} {\bibfnamefont {Y.}~\bibnamefont
  {Kim}}, \bibinfo {author} {\bibfnamefont {B.~J.}\ \bibnamefont {Wieder}},
  \bibinfo {author} {\bibfnamefont {C.~L.}\ \bibnamefont {Kane}}, \ and\
  \bibinfo {author} {\bibfnamefont {A.~M.}\ \bibnamefont {Rappe}},\ }\href
  {\doibase 10.1103/PhysRevLett.115.036806} {\bibfield  {journal} {\bibinfo
  {journal} {Phys. Rev. Lett.}\ }\textbf {\bibinfo {volume} {115}},\ \bibinfo
  {pages} {036806} (\bibinfo {year} {2015})}\BibitemShut {NoStop}%
\bibitem [{\citenamefont {Yu}\ \emph {et~al.}(2015)\citenamefont {Yu},
  \citenamefont {Weng}, \citenamefont {Fang}, \citenamefont {Dai},\ and\
  \citenamefont {Hu}}]{Yu15prl}%
  \BibitemOpen
  \bibfield  {author} {\bibinfo {author} {\bibfnamefont {R.}~\bibnamefont
  {Yu}}, \bibinfo {author} {\bibfnamefont {H.}~\bibnamefont {Weng}}, \bibinfo
  {author} {\bibfnamefont {Z.}~\bibnamefont {Fang}}, \bibinfo {author}
  {\bibfnamefont {X.}~\bibnamefont {Dai}}, \ and\ \bibinfo {author}
  {\bibfnamefont {X.}~\bibnamefont {Hu}},\ }\href {\doibase
  10.1103/PhysRevLett.115.036807} {\bibfield  {journal} {\bibinfo  {journal}
  {Phys. Rev. Lett.}\ }\textbf {\bibinfo {volume} {115}},\ \bibinfo {pages}
  {036807} (\bibinfo {year} {2015})}\BibitemShut {NoStop}%
\bibitem [{\citenamefont {Heikkil{\"a}}\ \emph {et~al.}(2011)\citenamefont
  {Heikkil{\"a}}, \citenamefont {Kopnin},\ and\ \citenamefont
  {Volovik}}]{Heikkila11jetp}%
  \BibitemOpen
  \bibfield  {author} {\bibinfo {author} {\bibfnamefont {T.~T.}\ \bibnamefont
  {Heikkil{\"a}}}, \bibinfo {author} {\bibfnamefont {N.~B.}\ \bibnamefont
  {Kopnin}}, \ and\ \bibinfo {author} {\bibfnamefont {G.~E.}\ \bibnamefont
  {Volovik}},\ }\href {\doibase 10.1134/S0021364011150045} {\bibfield
  {journal} {\bibinfo  {journal} {JETP Letters}\ }\textbf {\bibinfo {volume}
  {94}},\ \bibinfo {pages} {233} (\bibinfo {year} {2011})}\BibitemShut
  {NoStop}%
\bibitem [{\citenamefont {Weng}\ \emph {et~al.}(2015)\citenamefont {Weng},
  \citenamefont {Liang}, \citenamefont {Xu}, \citenamefont {Yu}, \citenamefont
  {Fang}, \citenamefont {Dai},\ and\ \citenamefont {Kawazoe}}]{Weng15prb}%
  \BibitemOpen
  \bibfield  {author} {\bibinfo {author} {\bibfnamefont {H.}~\bibnamefont
  {Weng}}, \bibinfo {author} {\bibfnamefont {Y.}~\bibnamefont {Liang}},
  \bibinfo {author} {\bibfnamefont {Q.}~\bibnamefont {Xu}}, \bibinfo {author}
  {\bibfnamefont {R.}~\bibnamefont {Yu}}, \bibinfo {author} {\bibfnamefont
  {Z.}~\bibnamefont {Fang}}, \bibinfo {author} {\bibfnamefont {X.}~\bibnamefont
  {Dai}}, \ and\ \bibinfo {author} {\bibfnamefont {Y.}~\bibnamefont
  {Kawazoe}},\ }\href {\doibase 10.1103/PhysRevB.92.045108} {\bibfield
  {journal} {\bibinfo  {journal} {Phys. Rev. B}\ }\textbf {\bibinfo {volume}
  {92}},\ \bibinfo {pages} {045108} (\bibinfo {year} {2015})}\BibitemShut
  {NoStop}%
\bibitem [{\citenamefont {Chen}\ \emph {et~al.}(2015)\citenamefont {Chen},
  \citenamefont {Xie}, \citenamefont {Yang}, \citenamefont {Pan}, \citenamefont
  {Zhang}, \citenamefont {Cohen},\ and\ \citenamefont {Zhang}}]{Chen15nl}%
  \BibitemOpen
  \bibfield  {author} {\bibinfo {author} {\bibfnamefont {Y.}~\bibnamefont
  {Chen}}, \bibinfo {author} {\bibfnamefont {Y.}~\bibnamefont {Xie}}, \bibinfo
  {author} {\bibfnamefont {S.~A.}\ \bibnamefont {Yang}}, \bibinfo {author}
  {\bibfnamefont {H.}~\bibnamefont {Pan}}, \bibinfo {author} {\bibfnamefont
  {F.}~\bibnamefont {Zhang}}, \bibinfo {author} {\bibfnamefont {M.~L.}\
  \bibnamefont {Cohen}}, \ and\ \bibinfo {author} {\bibfnamefont
  {S.}~\bibnamefont {Zhang}},\ }\href {\doibase 10.1021/acs.nanolett.5b02978}
  {\bibfield  {journal} {\bibinfo  {journal} {Nano Letters}\ }\textbf {\bibinfo
  {volume} {15}},\ \bibinfo {pages} {6974} (\bibinfo {year}
  {2015})}\BibitemShut {NoStop}%
\bibitem [{\citenamefont {Zeng}\ \emph {et~al.}(2015)\citenamefont {Zeng},
  \citenamefont {Fang}, \citenamefont {Chang}, \citenamefont {Chen},
  \citenamefont {Hsieh}, \citenamefont {Bansil}, \citenamefont {Lin},\ and\
  \citenamefont {Fu}}]{Zeng15arxiv}%
  \BibitemOpen
  \bibfield  {author} {\bibinfo {author} {\bibfnamefont {M.}~\bibnamefont
  {Zeng}}, \bibinfo {author} {\bibfnamefont {C.}~\bibnamefont {Fang}}, \bibinfo
  {author} {\bibfnamefont {G.}~\bibnamefont {Chang}}, \bibinfo {author}
  {\bibfnamefont {Y.-A.}\ \bibnamefont {Chen}}, \bibinfo {author}
  {\bibfnamefont {T.}~\bibnamefont {Hsieh}}, \bibinfo {author} {\bibfnamefont
  {A.}~\bibnamefont {Bansil}}, \bibinfo {author} {\bibfnamefont
  {H.}~\bibnamefont {Lin}}, \ and\ \bibinfo {author} {\bibfnamefont
  {L.}~\bibnamefont {Fu}},\ }\href {http://arxiv.org/abs/1504.03492} {\bibfield
   {journal} {\bibinfo  {journal} {arXiv:1504.03492 [cond-mat]}\ } (\bibinfo
  {year} {2015})},\ \bibinfo {note} {arXiv: 1504.03492}\BibitemShut {NoStop}%
\bibitem [{\citenamefont {Fang}\ \emph {et~al.}(2015)\citenamefont {Fang},
  \citenamefont {Chen}, \citenamefont {Kee},\ and\ \citenamefont
  {Fu}}]{Fang15prb}%
  \BibitemOpen
  \bibfield  {author} {\bibinfo {author} {\bibfnamefont {C.}~\bibnamefont
  {Fang}}, \bibinfo {author} {\bibfnamefont {Y.}~\bibnamefont {Chen}}, \bibinfo
  {author} {\bibfnamefont {H.-Y.}\ \bibnamefont {Kee}}, \ and\ \bibinfo
  {author} {\bibfnamefont {L.}~\bibnamefont {Fu}},\ }\href {\doibase
  10.1103/PhysRevB.92.081201} {\bibfield  {journal} {\bibinfo  {journal} {Phys.
  Rev. B}\ }\textbf {\bibinfo {volume} {92}},\ \bibinfo {pages} {081201}
  (\bibinfo {year} {2015})}\BibitemShut {NoStop}%
\bibitem [{\citenamefont {Mullen}\ \emph {et~al.}(2015)\citenamefont {Mullen},
  \citenamefont {Uchoa},\ and\ \citenamefont {Glatzhofer}}]{Mullen15prl}%
  \BibitemOpen
  \bibfield  {author} {\bibinfo {author} {\bibfnamefont {K.}~\bibnamefont
  {Mullen}}, \bibinfo {author} {\bibfnamefont {B.}~\bibnamefont {Uchoa}}, \
  and\ \bibinfo {author} {\bibfnamefont {D.~T.}\ \bibnamefont {Glatzhofer}},\
  }\href {\doibase 10.1103/PhysRevLett.115.026403} {\bibfield  {journal}
  {\bibinfo  {journal} {Phys. Rev. Lett.}\ }\textbf {\bibinfo {volume} {115}},\
  \bibinfo {pages} {026403} (\bibinfo {year} {2015})}\BibitemShut {NoStop}%
\bibitem [{\citenamefont {Yamakage}\ \emph {et~al.}(2016)\citenamefont
  {Yamakage}, \citenamefont {Yamakawa}, \citenamefont {Tanaka},\ and\
  \citenamefont {Okamoto}}]{Yamakage16jpsj}%
  \BibitemOpen
  \bibfield  {author} {\bibinfo {author} {\bibfnamefont {A.}~\bibnamefont
  {Yamakage}}, \bibinfo {author} {\bibfnamefont {Y.}~\bibnamefont {Yamakawa}},
  \bibinfo {author} {\bibfnamefont {Y.}~\bibnamefont {Tanaka}}, \ and\ \bibinfo
  {author} {\bibfnamefont {Y.}~\bibnamefont {Okamoto}},\ }\href {\doibase
  10.7566/jpsj.85.013708} {\bibfield  {journal} {\bibinfo  {journal} {Journal
  of the Physical Society of Japan}\ }\textbf {\bibinfo {volume} {85}},\
  \bibinfo {pages} {013708} (\bibinfo {year} {2016})}\BibitemShut {NoStop}%
\bibitem [{\citenamefont {Xie}\ \emph {et~al.}(2015)\citenamefont {Xie},
  \citenamefont {Schoop}, \citenamefont {Seibel}, \citenamefont {Gibson},
  \citenamefont {Xie},\ and\ \citenamefont {Cava}}]{Xie15aplm}%
  \BibitemOpen
  \bibfield  {author} {\bibinfo {author} {\bibfnamefont {L.~S.}\ \bibnamefont
  {Xie}}, \bibinfo {author} {\bibfnamefont {L.~M.}\ \bibnamefont {Schoop}},
  \bibinfo {author} {\bibfnamefont {E.~M.}\ \bibnamefont {Seibel}}, \bibinfo
  {author} {\bibfnamefont {Q.~D.}\ \bibnamefont {Gibson}}, \bibinfo {author}
  {\bibfnamefont {W.}~\bibnamefont {Xie}}, \ and\ \bibinfo {author}
  {\bibfnamefont {R.~J.}\ \bibnamefont {Cava}},\ }\href {\doibase
  10.1063/1.4926545} {\bibfield  {journal} {\bibinfo  {journal} {{APL}
  Materials}\ }\textbf {\bibinfo {volume} {3}},\ \bibinfo {pages} {083602}
  (\bibinfo {year} {2015})}\BibitemShut {NoStop}%
\bibitem [{\citenamefont {Chan}\ \emph {et~al.}(2016)\citenamefont {Chan},
  \citenamefont {Chiu}, \citenamefont {Chou},\ and\ \citenamefont
  {Schnyder}}]{Chan16prb}%
  \BibitemOpen
  \bibfield  {author} {\bibinfo {author} {\bibfnamefont {Y.-H.}\ \bibnamefont
  {Chan}}, \bibinfo {author} {\bibfnamefont {C.-K.}\ \bibnamefont {Chiu}},
  \bibinfo {author} {\bibfnamefont {M.~Y.}\ \bibnamefont {Chou}}, \ and\
  \bibinfo {author} {\bibfnamefont {A.~P.}\ \bibnamefont {Schnyder}},\ }\href
  {\doibase 10.1103/PhysRevB.93.205132} {\bibfield  {journal} {\bibinfo
  {journal} {Phys. Rev. B}\ }\textbf {\bibinfo {volume} {93}},\ \bibinfo
  {pages} {205132} (\bibinfo {year} {2016})}\BibitemShut {NoStop}%
\bibitem [{\citenamefont {Zhao}\ \emph {et~al.}(2016)\citenamefont {Zhao},
  \citenamefont {Yu}, \citenamefont {Weng},\ and\ \citenamefont
  {Fang}}]{Zhao16prb}%
  \BibitemOpen
  \bibfield  {author} {\bibinfo {author} {\bibfnamefont {J.}~\bibnamefont
  {Zhao}}, \bibinfo {author} {\bibfnamefont {R.}~\bibnamefont {Yu}}, \bibinfo
  {author} {\bibfnamefont {H.}~\bibnamefont {Weng}}, \ and\ \bibinfo {author}
  {\bibfnamefont {Z.}~\bibnamefont {Fang}},\ }\href {\doibase
  10.1103/PhysRevB.94.195104} {\bibfield  {journal} {\bibinfo  {journal} {Phys.
  Rev. B}\ }\textbf {\bibinfo {volume} {94}},\ \bibinfo {pages} {195104}
  (\bibinfo {year} {2016})}\BibitemShut {NoStop}%
\bibitem [{\citenamefont {Bian}\ \emph
  {et~al.}(2016{\natexlab{a}})\citenamefont {Bian}, \citenamefont {Chang},
  \citenamefont {Zheng}, \citenamefont {Velury}, \citenamefont {Xu},
  \citenamefont {Neupert}, \citenamefont {Chiu}, \citenamefont {Huang},
  \citenamefont {Sanchez}, \citenamefont {Belopolski}, \citenamefont
  {Alidoust}, \citenamefont {Chen}, \citenamefont {Chang}, \citenamefont
  {Bansil}, \citenamefont {Jeng}, \citenamefont {Lin},\ and\ \citenamefont
  {Hasan}}]{Bian16prb}%
  \BibitemOpen
  \bibfield  {author} {\bibinfo {author} {\bibfnamefont {G.}~\bibnamefont
  {Bian}}, \bibinfo {author} {\bibfnamefont {T.-R.}\ \bibnamefont {Chang}},
  \bibinfo {author} {\bibfnamefont {H.}~\bibnamefont {Zheng}}, \bibinfo
  {author} {\bibfnamefont {S.}~\bibnamefont {Velury}}, \bibinfo {author}
  {\bibfnamefont {S.-Y.}\ \bibnamefont {Xu}}, \bibinfo {author} {\bibfnamefont
  {T.}~\bibnamefont {Neupert}}, \bibinfo {author} {\bibfnamefont {C.-K.}\
  \bibnamefont {Chiu}}, \bibinfo {author} {\bibfnamefont {S.-M.}\ \bibnamefont
  {Huang}}, \bibinfo {author} {\bibfnamefont {D.~S.}\ \bibnamefont {Sanchez}},
  \bibinfo {author} {\bibfnamefont {I.}~\bibnamefont {Belopolski}}, \bibinfo
  {author} {\bibfnamefont {N.}~\bibnamefont {Alidoust}}, \bibinfo {author}
  {\bibfnamefont {P.-J.}\ \bibnamefont {Chen}}, \bibinfo {author}
  {\bibfnamefont {G.}~\bibnamefont {Chang}}, \bibinfo {author} {\bibfnamefont
  {A.}~\bibnamefont {Bansil}}, \bibinfo {author} {\bibfnamefont {H.-T.}\
  \bibnamefont {Jeng}}, \bibinfo {author} {\bibfnamefont {H.}~\bibnamefont
  {Lin}}, \ and\ \bibinfo {author} {\bibfnamefont {M.~Z.}\ \bibnamefont
  {Hasan}},\ }\href {\doibase 10.1103/PhysRevB.93.121113} {\bibfield  {journal}
  {\bibinfo  {journal} {Phys. Rev. B}\ }\textbf {\bibinfo {volume} {93}},\
  \bibinfo {pages} {121113} (\bibinfo {year} {2016}{\natexlab{a}})}\BibitemShut
  {NoStop}%
\bibitem [{\citenamefont {Bian}\ \emph
  {et~al.}(2016{\natexlab{b}})\citenamefont {Bian}, \citenamefont {Chang},
  \citenamefont {Sankar}, \citenamefont {Xu}, \citenamefont {Zheng},
  \citenamefont {Neupert}, \citenamefont {Chiu}, \citenamefont {Huang},
  \citenamefont {Chang}, \citenamefont {Belopolski}, \citenamefont {Sanchez},
  \citenamefont {Neupane}, \citenamefont {Alidoust}, \citenamefont {Liu},
  \citenamefont {Wang}, \citenamefont {Lee}, \citenamefont {Jeng},
  \citenamefont {Zhang}, \citenamefont {Yuan}, \citenamefont {Jia},
  \citenamefont {Bansil}, \citenamefont {Chou}, \citenamefont {Lin},\ and\
  \citenamefont {Hasan}}]{Bian16nc}%
  \BibitemOpen
  \bibfield  {author} {\bibinfo {author} {\bibfnamefont {G.}~\bibnamefont
  {Bian}}, \bibinfo {author} {\bibfnamefont {T.-R.}\ \bibnamefont {Chang}},
  \bibinfo {author} {\bibfnamefont {R.}~\bibnamefont {Sankar}}, \bibinfo
  {author} {\bibfnamefont {S.-Y.}\ \bibnamefont {Xu}}, \bibinfo {author}
  {\bibfnamefont {H.}~\bibnamefont {Zheng}}, \bibinfo {author} {\bibfnamefont
  {T.}~\bibnamefont {Neupert}}, \bibinfo {author} {\bibfnamefont {C.-K.}\
  \bibnamefont {Chiu}}, \bibinfo {author} {\bibfnamefont {S.-M.}\ \bibnamefont
  {Huang}}, \bibinfo {author} {\bibfnamefont {G.}~\bibnamefont {Chang}},
  \bibinfo {author} {\bibfnamefont {I.}~\bibnamefont {Belopolski}}, \bibinfo
  {author} {\bibfnamefont {D.~S.}\ \bibnamefont {Sanchez}}, \bibinfo {author}
  {\bibfnamefont {M.}~\bibnamefont {Neupane}}, \bibinfo {author} {\bibfnamefont
  {N.}~\bibnamefont {Alidoust}}, \bibinfo {author} {\bibfnamefont
  {C.}~\bibnamefont {Liu}}, \bibinfo {author} {\bibfnamefont {B.}~\bibnamefont
  {Wang}}, \bibinfo {author} {\bibfnamefont {C.-C.}\ \bibnamefont {Lee}},
  \bibinfo {author} {\bibfnamefont {H.-T.}\ \bibnamefont {Jeng}}, \bibinfo
  {author} {\bibfnamefont {C.}~\bibnamefont {Zhang}}, \bibinfo {author}
  {\bibfnamefont {Z.}~\bibnamefont {Yuan}}, \bibinfo {author} {\bibfnamefont
  {S.}~\bibnamefont {Jia}}, \bibinfo {author} {\bibfnamefont {A.}~\bibnamefont
  {Bansil}}, \bibinfo {author} {\bibfnamefont {F.}~\bibnamefont {Chou}},
  \bibinfo {author} {\bibfnamefont {H.}~\bibnamefont {Lin}}, \ and\ \bibinfo
  {author} {\bibfnamefont {M.~Z.}\ \bibnamefont {Hasan}},\ }\href {\doibase
  10.1038/ncomms10556} {\bibfield  {journal} {\bibinfo  {journal} {Nature
  Communications}\ }\textbf {\bibinfo {volume} {7}},\ \bibinfo {pages} {10556}
  (\bibinfo {year} {2016}{\natexlab{b}})}\BibitemShut {NoStop}%
\bibitem [{\citenamefont {Bzdu\v{s}ek}\ \emph {et~al.}(2016)\citenamefont
  {Bzdu\v{s}ek}, \citenamefont {Wu}, \citenamefont {R\"{u}egg}, \citenamefont
  {Sigrist},\ and\ \citenamefont {Soluyanov}}]{Bzdusek16nat}%
  \BibitemOpen
  \bibfield  {author} {\bibinfo {author} {\bibfnamefont {T.}~\bibnamefont
  {Bzdu\v{s}ek}}, \bibinfo {author} {\bibfnamefont {Q.}~\bibnamefont {Wu}},
  \bibinfo {author} {\bibfnamefont {A.}~\bibnamefont {R\"{u}egg}}, \bibinfo
  {author} {\bibfnamefont {M.}~\bibnamefont {Sigrist}}, \ and\ \bibinfo
  {author} {\bibfnamefont {A.~A.}\ \bibnamefont {Soluyanov}},\ }\href
  {http://dx.doi.org/10.1038/nature19099} {\bibfield  {journal} {\bibinfo
  {journal} {Nature}\ }\textbf {\bibinfo {volume} {538}},\ \bibinfo {pages}
  {75} (\bibinfo {year} {2016})},\ \bibinfo {note} {letter}\BibitemShut
  {NoStop}%
\bibitem [{\citenamefont {Chen}\ \emph {et~al.}(2017)\citenamefont {Chen},
  \citenamefont {Lu},\ and\ \citenamefont {Hou}}]{Chenwei17prb}%
  \BibitemOpen
  \bibfield  {author} {\bibinfo {author} {\bibfnamefont {W.}~\bibnamefont
  {Chen}}, \bibinfo {author} {\bibfnamefont {H.-Z.}\ \bibnamefont {Lu}}, \ and\
  \bibinfo {author} {\bibfnamefont {J.-M.}\ \bibnamefont {Hou}},\ }\href
  {\doibase 10.1103/PhysRevB.96.041102} {\bibfield  {journal} {\bibinfo
  {journal} {Phys. Rev. B}\ }\textbf {\bibinfo {volume} {96}},\ \bibinfo
  {pages} {041102} (\bibinfo {year} {2017})}\BibitemShut {NoStop}%
\bibitem [{\citenamefont {Yan}\ \emph {et~al.}(2017)\citenamefont {Yan},
  \citenamefont {Bi}, \citenamefont {Shen}, \citenamefont {Lu}, \citenamefont
  {Zhang},\ and\ \citenamefont {Wang}}]{Yan17prb}%
  \BibitemOpen
  \bibfield  {author} {\bibinfo {author} {\bibfnamefont {Z.}~\bibnamefont
  {Yan}}, \bibinfo {author} {\bibfnamefont {R.}~\bibnamefont {Bi}}, \bibinfo
  {author} {\bibfnamefont {H.}~\bibnamefont {Shen}}, \bibinfo {author}
  {\bibfnamefont {L.}~\bibnamefont {Lu}}, \bibinfo {author} {\bibfnamefont
  {S.-C.}\ \bibnamefont {Zhang}}, \ and\ \bibinfo {author} {\bibfnamefont
  {Z.}~\bibnamefont {Wang}},\ }\href {\doibase 10.1103/PhysRevB.96.041103}
  {\bibfield  {journal} {\bibinfo  {journal} {Phys. Rev. B}\ }\textbf {\bibinfo
  {volume} {96}},\ \bibinfo {pages} {041103} (\bibinfo {year}
  {2017})}\BibitemShut {NoStop}%
\bibitem [{\citenamefont {Ezawa}(2017)}]{Ezawa17prb}%
  \BibitemOpen
  \bibfield  {author} {\bibinfo {author} {\bibfnamefont {M.}~\bibnamefont
  {Ezawa}},\ }\href {\doibase 10.1103/PhysRevB.96.041202} {\bibfield  {journal}
  {\bibinfo  {journal} {Phys. Rev. B}\ }\textbf {\bibinfo {volume} {96}},\
  \bibinfo {pages} {041202} (\bibinfo {year} {2017})}\BibitemShut {NoStop}%
\bibitem [{\citenamefont {Schoop}\ \emph {et~al.}(2016)\citenamefont {Schoop},
  \citenamefont {Ali}, \citenamefont {Stra{\ss}er}, \citenamefont {Topp},
  \citenamefont {Varykhalov}, \citenamefont {Marchenko}, \citenamefont
  {Duppel}, \citenamefont {Parkin}, \citenamefont {Lotsch},\ and\ \citenamefont
  {Ast}}]{Schoop16nc}%
  \BibitemOpen
  \bibfield  {author} {\bibinfo {author} {\bibfnamefont {L.~M.}\ \bibnamefont
  {Schoop}}, \bibinfo {author} {\bibfnamefont {M.~N.}\ \bibnamefont {Ali}},
  \bibinfo {author} {\bibfnamefont {C.}~\bibnamefont {Stra{\ss}er}}, \bibinfo
  {author} {\bibfnamefont {A.}~\bibnamefont {Topp}}, \bibinfo {author}
  {\bibfnamefont {A.}~\bibnamefont {Varykhalov}}, \bibinfo {author}
  {\bibfnamefont {D.}~\bibnamefont {Marchenko}}, \bibinfo {author}
  {\bibfnamefont {V.}~\bibnamefont {Duppel}}, \bibinfo {author} {\bibfnamefont
  {S.~S.~P.}\ \bibnamefont {Parkin}}, \bibinfo {author} {\bibfnamefont {B.~V.}\
  \bibnamefont {Lotsch}}, \ and\ \bibinfo {author} {\bibfnamefont {C.~R.}\
  \bibnamefont {Ast}},\ }\href {http://dx.doi.org/10.1038/ncomms11696} {\
  \textbf {\bibinfo {volume} {7}},\ \bibinfo {pages} {11696} (\bibinfo {year}
  {2016})}\BibitemShut {NoStop}%
\bibitem [{\citenamefont {Neupane}\ \emph {et~al.}(2016)\citenamefont
  {Neupane}, \citenamefont {Belopolski}, \citenamefont {Hosen}, \citenamefont
  {Sanchez}, \citenamefont {Sankar}, \citenamefont {Szlawska}, \citenamefont
  {Xu}, \citenamefont {Dimitri}, \citenamefont {Dhakal}, \citenamefont
  {Maldonado}, \citenamefont {Oppeneer}, \citenamefont {Kaczorowski},
  \citenamefont {Chou}, \citenamefont {Hasan},\ and\ \citenamefont
  {Durakiewicz}}]{Neupane16prb}%
  \BibitemOpen
  \bibfield  {author} {\bibinfo {author} {\bibfnamefont {M.}~\bibnamefont
  {Neupane}}, \bibinfo {author} {\bibfnamefont {I.}~\bibnamefont {Belopolski}},
  \bibinfo {author} {\bibfnamefont {M.~M.}\ \bibnamefont {Hosen}}, \bibinfo
  {author} {\bibfnamefont {D.~S.}\ \bibnamefont {Sanchez}}, \bibinfo {author}
  {\bibfnamefont {R.}~\bibnamefont {Sankar}}, \bibinfo {author} {\bibfnamefont
  {M.}~\bibnamefont {Szlawska}}, \bibinfo {author} {\bibfnamefont {S.-Y.}\
  \bibnamefont {Xu}}, \bibinfo {author} {\bibfnamefont {K.}~\bibnamefont
  {Dimitri}}, \bibinfo {author} {\bibfnamefont {N.}~\bibnamefont {Dhakal}},
  \bibinfo {author} {\bibfnamefont {P.}~\bibnamefont {Maldonado}}, \bibinfo
  {author} {\bibfnamefont {P.~M.}\ \bibnamefont {Oppeneer}}, \bibinfo {author}
  {\bibfnamefont {D.}~\bibnamefont {Kaczorowski}}, \bibinfo {author}
  {\bibfnamefont {F.}~\bibnamefont {Chou}}, \bibinfo {author} {\bibfnamefont
  {M.~Z.}\ \bibnamefont {Hasan}}, \ and\ \bibinfo {author} {\bibfnamefont
  {T.}~\bibnamefont {Durakiewicz}},\ }\href {\doibase
  10.1103/PhysRevB.93.201104} {\bibfield  {journal} {\bibinfo  {journal} {Phys.
  Rev. B}\ }\textbf {\bibinfo {volume} {93}},\ \bibinfo {pages} {201104}
  (\bibinfo {year} {2016})}\BibitemShut {NoStop}%
\bibitem [{\citenamefont {Topp}\ \emph {et~al.}(2016)\citenamefont {Topp},
  \citenamefont {Lippmann}, \citenamefont {Varykhalov}, \citenamefont {Duppel},
  \citenamefont {Lotsch}, \citenamefont {Ast},\ and\ \citenamefont
  {Schoop}}]{Andreas16njp}%
  \BibitemOpen
  \bibfield  {author} {\bibinfo {author} {\bibfnamefont {A.}~\bibnamefont
  {Topp}}, \bibinfo {author} {\bibfnamefont {J.~M.}\ \bibnamefont {Lippmann}},
  \bibinfo {author} {\bibfnamefont {A.}~\bibnamefont {Varykhalov}}, \bibinfo
  {author} {\bibfnamefont {V.}~\bibnamefont {Duppel}}, \bibinfo {author}
  {\bibfnamefont {B.~V.}\ \bibnamefont {Lotsch}}, \bibinfo {author}
  {\bibfnamefont {C.~R.}\ \bibnamefont {Ast}}, \ and\ \bibinfo {author}
  {\bibfnamefont {L.~M.}\ \bibnamefont {Schoop}},\ }\href
  {http://stacks.iop.org/1367-2630/18/i=12/a=125014} {\bibfield  {journal}
  {\bibinfo  {journal} {New Journal of Physics}\ }\textbf {\bibinfo {volume}
  {18}},\ \bibinfo {pages} {125014} (\bibinfo {year} {2016})}\BibitemShut
  {NoStop}%
\bibitem [{\citenamefont {Takane}\ \emph {et~al.}(2016)\citenamefont {Takane},
  \citenamefont {Wang}, \citenamefont {Souma}, \citenamefont {Nakayama},
  \citenamefont {Trang}, \citenamefont {Sato}, \citenamefont {Takahashi},\ and\
  \citenamefont {Ando}}]{Takane16prb}%
  \BibitemOpen
  \bibfield  {author} {\bibinfo {author} {\bibfnamefont {D.}~\bibnamefont
  {Takane}}, \bibinfo {author} {\bibfnamefont {Z.}~\bibnamefont {Wang}},
  \bibinfo {author} {\bibfnamefont {S.}~\bibnamefont {Souma}}, \bibinfo
  {author} {\bibfnamefont {K.}~\bibnamefont {Nakayama}}, \bibinfo {author}
  {\bibfnamefont {C.~X.}\ \bibnamefont {Trang}}, \bibinfo {author}
  {\bibfnamefont {T.}~\bibnamefont {Sato}}, \bibinfo {author} {\bibfnamefont
  {T.}~\bibnamefont {Takahashi}}, \ and\ \bibinfo {author} {\bibfnamefont
  {Y.}~\bibnamefont {Ando}},\ }\href {\doibase 10.1103/PhysRevB.94.121108}
  {\bibfield  {journal} {\bibinfo  {journal} {Phys. Rev. B}\ }\textbf {\bibinfo
  {volume} {94}},\ \bibinfo {pages} {121108} (\bibinfo {year}
  {2016})}\BibitemShut {NoStop}%
\bibitem [{\citenamefont {Hu}\ \emph {et~al.}(2016)\citenamefont {Hu},
  \citenamefont {Tang}, \citenamefont {Liu}, \citenamefont {Liu}, \citenamefont
  {Zhu}, \citenamefont {Graf}, \citenamefont {Myhro}, \citenamefont {Tran},
  \citenamefont {Lau}, \citenamefont {Wei},\ and\ \citenamefont
  {Mao}}]{Hu16prl}%
  \BibitemOpen
  \bibfield  {author} {\bibinfo {author} {\bibfnamefont {J.}~\bibnamefont
  {Hu}}, \bibinfo {author} {\bibfnamefont {Z.}~\bibnamefont {Tang}}, \bibinfo
  {author} {\bibfnamefont {J.}~\bibnamefont {Liu}}, \bibinfo {author}
  {\bibfnamefont {X.}~\bibnamefont {Liu}}, \bibinfo {author} {\bibfnamefont
  {Y.}~\bibnamefont {Zhu}}, \bibinfo {author} {\bibfnamefont {D.}~\bibnamefont
  {Graf}}, \bibinfo {author} {\bibfnamefont {K.}~\bibnamefont {Myhro}},
  \bibinfo {author} {\bibfnamefont {S.}~\bibnamefont {Tran}}, \bibinfo {author}
  {\bibfnamefont {C.~N.}\ \bibnamefont {Lau}}, \bibinfo {author} {\bibfnamefont
  {J.}~\bibnamefont {Wei}}, \ and\ \bibinfo {author} {\bibfnamefont
  {Z.}~\bibnamefont {Mao}},\ }\href {\doibase 10.1103/PhysRevLett.117.016602}
  {\bibfield  {journal} {\bibinfo  {journal} {Phys. Rev. Lett.}\ }\textbf
  {\bibinfo {volume} {117}},\ \bibinfo {pages} {016602} (\bibinfo {year}
  {2016})}\BibitemShut {NoStop}%
\bibitem [{\citenamefont {Hu}\ \emph {et~al.}(2017)\citenamefont {Hu},
  \citenamefont {Zhu}, \citenamefont {Graf}, \citenamefont {Tang},
  \citenamefont {Liu},\ and\ \citenamefont {Mao}}]{Hu17prb}%
  \BibitemOpen
  \bibfield  {author} {\bibinfo {author} {\bibfnamefont {J.}~\bibnamefont
  {Hu}}, \bibinfo {author} {\bibfnamefont {Y.~L.}\ \bibnamefont {Zhu}},
  \bibinfo {author} {\bibfnamefont {D.}~\bibnamefont {Graf}}, \bibinfo {author}
  {\bibfnamefont {Z.~J.}\ \bibnamefont {Tang}}, \bibinfo {author}
  {\bibfnamefont {J.~Y.}\ \bibnamefont {Liu}}, \ and\ \bibinfo {author}
  {\bibfnamefont {Z.~Q.}\ \bibnamefont {Mao}},\ }\href {\doibase
  10.1103/PhysRevB.95.205134} {\bibfield  {journal} {\bibinfo  {journal} {Phys.
  Rev. B}\ }\textbf {\bibinfo {volume} {95}},\ \bibinfo {pages} {205134}
  (\bibinfo {year} {2017})}\BibitemShut {NoStop}%
\bibitem [{\citenamefont {Kumar}\ \emph {et~al.}(2017)\citenamefont {Kumar},
  \citenamefont {Manna}, \citenamefont {Qi}, \citenamefont {Wu}, \citenamefont
  {Wang}, \citenamefont {Yan}, \citenamefont {Felser},\ and\ \citenamefont
  {Shekhar}}]{Kumar17prb}%
  \BibitemOpen
  \bibfield  {author} {\bibinfo {author} {\bibfnamefont {N.}~\bibnamefont
  {Kumar}}, \bibinfo {author} {\bibfnamefont {K.}~\bibnamefont {Manna}},
  \bibinfo {author} {\bibfnamefont {Y.}~\bibnamefont {Qi}}, \bibinfo {author}
  {\bibfnamefont {S.-C.}\ \bibnamefont {Wu}}, \bibinfo {author} {\bibfnamefont
  {L.}~\bibnamefont {Wang}}, \bibinfo {author} {\bibfnamefont {B.}~\bibnamefont
  {Yan}}, \bibinfo {author} {\bibfnamefont {C.}~\bibnamefont {Felser}}, \ and\
  \bibinfo {author} {\bibfnamefont {C.}~\bibnamefont {Shekhar}},\ }\href
  {\doibase 10.1103/PhysRevB.95.121109} {\bibfield  {journal} {\bibinfo
  {journal} {Phys. Rev. B}\ }\textbf {\bibinfo {volume} {95}},\ \bibinfo
  {pages} {121109} (\bibinfo {year} {2017})}\BibitemShut {NoStop}%
\bibitem [{\citenamefont {Pan}\ \emph {et~al.}(2017)\citenamefont {Pan},
  \citenamefont {Tong}, \citenamefont {Yu}, \citenamefont {Wang}, \citenamefont
  {Fu}, \citenamefont {Zhang}, \citenamefont {Wu}, \citenamefont {Wan},
  \citenamefont {Zhang}, \citenamefont {Wang},\ and\ \citenamefont
  {Song}}]{Pan17arxiv}%
  \BibitemOpen
  \bibfield  {author} {\bibinfo {author} {\bibfnamefont {H.}~\bibnamefont
  {Pan}}, \bibinfo {author} {\bibfnamefont {B.}~\bibnamefont {Tong}}, \bibinfo
  {author} {\bibfnamefont {J.}~\bibnamefont {Yu}}, \bibinfo {author}
  {\bibfnamefont {J.}~\bibnamefont {Wang}}, \bibinfo {author} {\bibfnamefont
  {D.}~\bibnamefont {Fu}}, \bibinfo {author} {\bibfnamefont {S.}~\bibnamefont
  {Zhang}}, \bibinfo {author} {\bibfnamefont {B.}~\bibnamefont {Wu}}, \bibinfo
  {author} {\bibfnamefont {X.}~\bibnamefont {Wan}}, \bibinfo {author}
  {\bibfnamefont {C.}~\bibnamefont {Zhang}}, \bibinfo {author} {\bibfnamefont
  {X.}~\bibnamefont {Wang}}, \ and\ \bibinfo {author} {\bibfnamefont
  {F.}~\bibnamefont {Song}},\ }\href {http://arxiv.org/abs/1708.02779}
  {\bibfield  {journal} {\bibinfo  {journal} {arXiv:1708.02779 [cond-mat]}\ }
  (\bibinfo {year} {2017})},\ \bibinfo {note} {arXiv: 1708.02779}\BibitemShut
  {NoStop}%
\bibitem [{\citenamefont {Li}\ \emph {et~al.}(2018{\natexlab{a}})\citenamefont
  {Li}, \citenamefont {Wang}, \citenamefont {Wan}, \citenamefont {Wan},
  \citenamefont {Lu},\ and\ \citenamefont {Xie}}]{Li18prl}%
  \BibitemOpen
  \bibfield  {author} {\bibinfo {author} {\bibfnamefont {C.}~\bibnamefont
  {Li}}, \bibinfo {author} {\bibfnamefont {C.~M.}\ \bibnamefont {Wang}},
  \bibinfo {author} {\bibfnamefont {B.}~\bibnamefont {Wan}}, \bibinfo {author}
  {\bibfnamefont {X.}~\bibnamefont {Wan}}, \bibinfo {author} {\bibfnamefont
  {H.-Z.}\ \bibnamefont {Lu}}, \ and\ \bibinfo {author} {\bibfnamefont {X.~C.}\
  \bibnamefont {Xie}},\ }\href {\doibase 10.1103/PhysRevLett.120.146602}
  {\bibfield  {journal} {\bibinfo  {journal} {Phys. Rev. Lett.}\ }\textbf
  {\bibinfo {volume} {120}},\ \bibinfo {pages} {146602} (\bibinfo {year}
  {2018}{\natexlab{a}})}\BibitemShut {NoStop}%
\bibitem [{\citenamefont {Chen}\ \emph {et~al.}(2018)\citenamefont {Chen},
  \citenamefont {Luo}, \citenamefont {Li},\ and\ \citenamefont
  {Zilberberg}}]{Chen18prl}%
  \BibitemOpen
  \bibfield  {author} {\bibinfo {author} {\bibfnamefont {W.}~\bibnamefont
  {Chen}}, \bibinfo {author} {\bibfnamefont {K.}~\bibnamefont {Luo}}, \bibinfo
  {author} {\bibfnamefont {L.}~\bibnamefont {Li}}, \ and\ \bibinfo {author}
  {\bibfnamefont {O.}~\bibnamefont {Zilberberg}},\ }\href {\doibase
  10.1103/PhysRevLett.121.166802} {\bibfield  {journal} {\bibinfo  {journal}
  {Phys. Rev. Lett.}\ }\textbf {\bibinfo {volume} {121}},\ \bibinfo {pages}
  {166802} (\bibinfo {year} {2018})}\BibitemShut {NoStop}%
\bibitem [{\citenamefont {Takane}\ \emph {et~al.}(2018)\citenamefont {Takane},
  \citenamefont {Nakayama}, \citenamefont {Souma}, \citenamefont {Wada},
  \citenamefont {Okamoto}, \citenamefont {Takenaka}, \citenamefont {Yamakawa},
  \citenamefont {Yamakage}, \citenamefont {Mitsuhashi}, \citenamefont {Horiba}
  \emph {et~al.}}]{Takane18npj}%
  \BibitemOpen
  \bibfield  {author} {\bibinfo {author} {\bibfnamefont {D.}~\bibnamefont
  {Takane}}, \bibinfo {author} {\bibfnamefont {K.}~\bibnamefont {Nakayama}},
  \bibinfo {author} {\bibfnamefont {S.}~\bibnamefont {Souma}}, \bibinfo
  {author} {\bibfnamefont {T.}~\bibnamefont {Wada}}, \bibinfo {author}
  {\bibfnamefont {Y.}~\bibnamefont {Okamoto}}, \bibinfo {author} {\bibfnamefont
  {K.}~\bibnamefont {Takenaka}}, \bibinfo {author} {\bibfnamefont
  {Y.}~\bibnamefont {Yamakawa}}, \bibinfo {author} {\bibfnamefont
  {A.}~\bibnamefont {Yamakage}}, \bibinfo {author} {\bibfnamefont
  {T.}~\bibnamefont {Mitsuhashi}}, \bibinfo {author} {\bibfnamefont
  {K.}~\bibnamefont {Horiba}},  \emph {et~al.},\ }\href@noop {} {\bibfield
  {journal} {\bibinfo  {journal} {npj Quantum Materials}\ }\textbf {\bibinfo
  {volume} {3}},\ \bibinfo {pages} {1} (\bibinfo {year} {2018})}\BibitemShut
  {NoStop}%
\bibitem [{\citenamefont {Syzranov}\ and\ \citenamefont
  {Skinner}(2017)}]{Syzranov17prb}%
  \BibitemOpen
  \bibfield  {author} {\bibinfo {author} {\bibfnamefont {S.~V.}\ \bibnamefont
  {Syzranov}}\ and\ \bibinfo {author} {\bibfnamefont {B.}~\bibnamefont
  {Skinner}},\ }\href {\doibase 10.1103/PhysRevB.96.161105} {\bibfield
  {journal} {\bibinfo  {journal} {Phys. Rev. B}\ }\textbf {\bibinfo {volume}
  {96}},\ \bibinfo {pages} {161105} (\bibinfo {year} {2017})}\BibitemShut
  {NoStop}%
\bibitem [{sup()}]{supmat}%
  \BibitemOpen
  \href@noop {} {}\bibinfo {note} {See Supplemental Material at ... for the
  derivation of the elastic scattering time, vertex correction to velocity,
  Drude conductivity, Cooperon formula, quantum correction to the conductivity,
  Magnetoconductivity in both limiting cases, and the discussion on the effects
  of temperature, spin-orbit coupling, electron-electron interaction, tilted
  nodal line and Berry curvature, which includes Refs.
  \cite{Akkermans07,Datta97book,Altshuler85,Li17prb,Ahn17prl,Huang18prb,Rui18prb,Burkov18prb,Son13prb,Dai17prl}}\BibitemShut
  {NoStop}%
\bibitem [{gen()}]{generality}%
  \BibitemOpen
  \href@noop {} {}\bibinfo {note} {The Fermi surface of the nodal-line
  semimetal is anisotropic and generates an anisotropic conductivity.
  Nevertheless, the qualitative difference between the calculated WL and WAL
  corrections do not change with the transport direction.}\BibitemShut {Stop}%
\bibitem [{\citenamefont {Chalaev}\ and\ \citenamefont
  {Loss}(2005)}]{Chalaev05prb}%
  \BibitemOpen
  \bibfield  {author} {\bibinfo {author} {\bibfnamefont {O.}~\bibnamefont
  {Chalaev}}\ and\ \bibinfo {author} {\bibfnamefont {D.}~\bibnamefont {Loss}},\
  }\href {\doibase 10.1103/PhysRevB.71.245318} {\bibfield  {journal} {\bibinfo
  {journal} {Phys. Rev. B}\ }\textbf {\bibinfo {volume} {71}},\ \bibinfo
  {pages} {245318} (\bibinfo {year} {2005})}\BibitemShut {NoStop}%
\bibitem [{tim()}]{time}%
  \BibitemOpen
  \href@noop {} {}\bibinfo {note} {The elastic scattering time is
  $\tau_e=\hbar/(\Delta\rho_0\gamma)$ in the LR limit, so that $\sigma_z^L$ is
  a function of $\Delta$.}\BibitemShut {Stop}%
\bibitem [{\citenamefont {Cao}\ \emph {et~al.}(2012)\citenamefont {Cao},
  \citenamefont {Tian}, \citenamefont {Miotkowski}, \citenamefont {Shen},
  \citenamefont {Hu}, \citenamefont {Qiao},\ and\ \citenamefont
  {Chen}}]{Cao12prl}%
  \BibitemOpen
  \bibfield  {author} {\bibinfo {author} {\bibfnamefont {H.}~\bibnamefont
  {Cao}}, \bibinfo {author} {\bibfnamefont {J.}~\bibnamefont {Tian}}, \bibinfo
  {author} {\bibfnamefont {I.}~\bibnamefont {Miotkowski}}, \bibinfo {author}
  {\bibfnamefont {T.}~\bibnamefont {Shen}}, \bibinfo {author} {\bibfnamefont
  {J.}~\bibnamefont {Hu}}, \bibinfo {author} {\bibfnamefont {S.}~\bibnamefont
  {Qiao}}, \ and\ \bibinfo {author} {\bibfnamefont {Y.~P.}\ \bibnamefont
  {Chen}},\ }\href {\doibase 10.1103/PhysRevLett.108.216803} {\bibfield
  {journal} {\bibinfo  {journal} {Phys. Rev. Lett.}\ }\textbf {\bibinfo
  {volume} {108}},\ \bibinfo {pages} {216803} (\bibinfo {year}
  {2012})}\BibitemShut {NoStop}%
\bibitem [{\citenamefont {Pan}\ \emph {et~al.}(2018)\citenamefont {Pan},
  \citenamefont {Tong}, \citenamefont {Yu}, \citenamefont {Wang}, \citenamefont
  {Fu}, \citenamefont {Zhang}, \citenamefont {Wu}, \citenamefont {Wan},
  \citenamefont {Zhang}, \citenamefont {Wang} \emph {et~al.}}]{Pan18sr}%
  \BibitemOpen
  \bibfield  {author} {\bibinfo {author} {\bibfnamefont {H.}~\bibnamefont
  {Pan}}, \bibinfo {author} {\bibfnamefont {B.}~\bibnamefont {Tong}}, \bibinfo
  {author} {\bibfnamefont {J.}~\bibnamefont {Yu}}, \bibinfo {author}
  {\bibfnamefont {J.}~\bibnamefont {Wang}}, \bibinfo {author} {\bibfnamefont
  {D.}~\bibnamefont {Fu}}, \bibinfo {author} {\bibfnamefont {S.}~\bibnamefont
  {Zhang}}, \bibinfo {author} {\bibfnamefont {B.}~\bibnamefont {Wu}}, \bibinfo
  {author} {\bibfnamefont {X.}~\bibnamefont {Wan}}, \bibinfo {author}
  {\bibfnamefont {C.}~\bibnamefont {Zhang}}, \bibinfo {author} {\bibfnamefont
  {X.}~\bibnamefont {Wang}},  \emph {et~al.},\ }\href@noop {} {\bibfield
  {journal} {\bibinfo  {journal} {Scientific Reports (Nature Publisher Group)}\
  }\textbf {\bibinfo {volume} {8}},\ \bibinfo {pages} {1} (\bibinfo {year}
  {2018})}\BibitemShut {NoStop}%
\bibitem [{\citenamefont {Wang}\ \emph {et~al.}(2016)\citenamefont {Wang},
  \citenamefont {Pan}, \citenamefont {Gao}, \citenamefont {Yu}, \citenamefont
  {Jiang}, \citenamefont {Zhang}, \citenamefont {Zuo}, \citenamefont {Zhang},
  \citenamefont {Wei}, \citenamefont {Niu} \emph {et~al.}}]{Wang16aem}%
  \BibitemOpen
  \bibfield  {author} {\bibinfo {author} {\bibfnamefont {X.}~\bibnamefont
  {Wang}}, \bibinfo {author} {\bibfnamefont {X.}~\bibnamefont {Pan}}, \bibinfo
  {author} {\bibfnamefont {M.}~\bibnamefont {Gao}}, \bibinfo {author}
  {\bibfnamefont {J.}~\bibnamefont {Yu}}, \bibinfo {author} {\bibfnamefont
  {J.}~\bibnamefont {Jiang}}, \bibinfo {author} {\bibfnamefont
  {J.}~\bibnamefont {Zhang}}, \bibinfo {author} {\bibfnamefont
  {H.}~\bibnamefont {Zuo}}, \bibinfo {author} {\bibfnamefont {M.}~\bibnamefont
  {Zhang}}, \bibinfo {author} {\bibfnamefont {Z.}~\bibnamefont {Wei}}, \bibinfo
  {author} {\bibfnamefont {W.}~\bibnamefont {Niu}},  \emph {et~al.},\
  }\href@noop {} {\bibfield  {journal} {\bibinfo  {journal} {Advanced
  Electronic Materials}\ }\textbf {\bibinfo {volume} {2}},\ \bibinfo {pages}
  {1600228} (\bibinfo {year} {2016})}\BibitemShut {NoStop}%
\bibitem [{\citenamefont {Lv}\ \emph {et~al.}(2016)\citenamefont {Lv},
  \citenamefont {Zhang}, \citenamefont {Li}, \citenamefont {Yao}, \citenamefont
  {Chen}, \citenamefont {Zhou}, \citenamefont {Zhang}, \citenamefont {Lu},\
  and\ \citenamefont {Chen}}]{Lv16apl}%
  \BibitemOpen
  \bibfield  {author} {\bibinfo {author} {\bibfnamefont {Y.-Y.}\ \bibnamefont
  {Lv}}, \bibinfo {author} {\bibfnamefont {B.-B.}\ \bibnamefont {Zhang}},
  \bibinfo {author} {\bibfnamefont {X.}~\bibnamefont {Li}}, \bibinfo {author}
  {\bibfnamefont {S.-H.}\ \bibnamefont {Yao}}, \bibinfo {author} {\bibfnamefont
  {Y.}~\bibnamefont {Chen}}, \bibinfo {author} {\bibfnamefont {J.}~\bibnamefont
  {Zhou}}, \bibinfo {author} {\bibfnamefont {S.-T.}\ \bibnamefont {Zhang}},
  \bibinfo {author} {\bibfnamefont {M.-H.}\ \bibnamefont {Lu}}, \ and\ \bibinfo
  {author} {\bibfnamefont {Y.-F.}\ \bibnamefont {Chen}},\ }\href@noop {}
  {\bibfield  {journal} {\bibinfo  {journal} {Applied Physics Letters}\
  }\textbf {\bibinfo {volume} {108}},\ \bibinfo {pages} {244101} (\bibinfo
  {year} {2016})}\BibitemShut {NoStop}%
\bibitem [{\citenamefont {Ali}\ \emph {et~al.}(2016)\citenamefont {Ali},
  \citenamefont {Schoop}, \citenamefont {Garg}, \citenamefont {Lippmann},
  \citenamefont {Lara}, \citenamefont {Lotsch},\ and\ \citenamefont
  {Parkin}}]{Ali16sa}%
  \BibitemOpen
  \bibfield  {author} {\bibinfo {author} {\bibfnamefont {M.~N.}\ \bibnamefont
  {Ali}}, \bibinfo {author} {\bibfnamefont {L.~M.}\ \bibnamefont {Schoop}},
  \bibinfo {author} {\bibfnamefont {C.}~\bibnamefont {Garg}}, \bibinfo {author}
  {\bibfnamefont {J.~M.}\ \bibnamefont {Lippmann}}, \bibinfo {author}
  {\bibfnamefont {E.}~\bibnamefont {Lara}}, \bibinfo {author} {\bibfnamefont
  {B.}~\bibnamefont {Lotsch}}, \ and\ \bibinfo {author} {\bibfnamefont {S.~S.}\
  \bibnamefont {Parkin}},\ }\href@noop {} {\bibfield  {journal} {\bibinfo
  {journal} {Science advances}\ }\textbf {\bibinfo {volume} {2}},\ \bibinfo
  {pages} {e1601742} (\bibinfo {year} {2016})}\BibitemShut {NoStop}%
\bibitem [{\citenamefont {Zhang}\ \emph {et~al.}(2018)\citenamefont {Zhang},
  \citenamefont {Gao}, \citenamefont {Zhang}, \citenamefont {Wang},
  \citenamefont {Zhang}, \citenamefont {Zhang}, \citenamefont {Niu},
  \citenamefont {Zhang},\ and\ \citenamefont {Xu}}]{Zhang18fp}%
  \BibitemOpen
  \bibfield  {author} {\bibinfo {author} {\bibfnamefont {J.}~\bibnamefont
  {Zhang}}, \bibinfo {author} {\bibfnamefont {M.}~\bibnamefont {Gao}}, \bibinfo
  {author} {\bibfnamefont {J.}~\bibnamefont {Zhang}}, \bibinfo {author}
  {\bibfnamefont {X.}~\bibnamefont {Wang}}, \bibinfo {author} {\bibfnamefont
  {X.}~\bibnamefont {Zhang}}, \bibinfo {author} {\bibfnamefont
  {M.}~\bibnamefont {Zhang}}, \bibinfo {author} {\bibfnamefont
  {W.}~\bibnamefont {Niu}}, \bibinfo {author} {\bibfnamefont {R.}~\bibnamefont
  {Zhang}}, \ and\ \bibinfo {author} {\bibfnamefont {Y.}~\bibnamefont {Xu}},\
  }\href@noop {} {\bibfield  {journal} {\bibinfo  {journal} {Frontiers of
  Physics}\ }\textbf {\bibinfo {volume} {13}},\ \bibinfo {pages} {137201}
  (\bibinfo {year} {2018})}\BibitemShut {NoStop}%
\bibitem [{\citenamefont {Li}\ \emph {et~al.}(2018{\natexlab{b}})\citenamefont
  {Li}, \citenamefont {Guo}, \citenamefont {Fu}, \citenamefont {Pan},
  \citenamefont {Wang}, \citenamefont {Ran}, \citenamefont {Bao}, \citenamefont
  {Ma}, \citenamefont {Cai}, \citenamefont {Wang} \emph {et~al.}}]{Li18sb}%
  \BibitemOpen
  \bibfield  {author} {\bibinfo {author} {\bibfnamefont {S.}~\bibnamefont
  {Li}}, \bibinfo {author} {\bibfnamefont {Z.}~\bibnamefont {Guo}}, \bibinfo
  {author} {\bibfnamefont {D.}~\bibnamefont {Fu}}, \bibinfo {author}
  {\bibfnamefont {X.-C.}\ \bibnamefont {Pan}}, \bibinfo {author} {\bibfnamefont
  {J.}~\bibnamefont {Wang}}, \bibinfo {author} {\bibfnamefont {K.}~\bibnamefont
  {Ran}}, \bibinfo {author} {\bibfnamefont {S.}~\bibnamefont {Bao}}, \bibinfo
  {author} {\bibfnamefont {Z.}~\bibnamefont {Ma}}, \bibinfo {author}
  {\bibfnamefont {Z.}~\bibnamefont {Cai}}, \bibinfo {author} {\bibfnamefont
  {R.}~\bibnamefont {Wang}},  \emph {et~al.},\ }\href@noop {} {\bibfield
  {journal} {\bibinfo  {journal} {Science Bulletin}\ }\textbf {\bibinfo
  {volume} {63}},\ \bibinfo {pages} {535} (\bibinfo {year}
  {2018}{\natexlab{b}})}\BibitemShut {NoStop}%
\bibitem [{\citenamefont {An}\ \emph {et~al.}(2019)\citenamefont {An},
  \citenamefont {Zhu}, \citenamefont {Gao}, \citenamefont {Wu}, \citenamefont
  {Ning},\ and\ \citenamefont {Tian}}]{An19prb}%
  \BibitemOpen
  \bibfield  {author} {\bibinfo {author} {\bibfnamefont {L.}~\bibnamefont
  {An}}, \bibinfo {author} {\bibfnamefont {X.}~\bibnamefont {Zhu}}, \bibinfo
  {author} {\bibfnamefont {W.}~\bibnamefont {Gao}}, \bibinfo {author}
  {\bibfnamefont {M.}~\bibnamefont {Wu}}, \bibinfo {author} {\bibfnamefont
  {W.}~\bibnamefont {Ning}}, \ and\ \bibinfo {author} {\bibfnamefont
  {M.}~\bibnamefont {Tian}},\ }\href {\doibase 10.1103/PhysRevB.99.045143}
  {\bibfield  {journal} {\bibinfo  {journal} {Phys. Rev. B}\ }\textbf {\bibinfo
  {volume} {99}},\ \bibinfo {pages} {045143} (\bibinfo {year}
  {2019})}\BibitemShut {NoStop}%
\bibitem [{\citenamefont {Datta}(1997)}]{Datta97book}%
  \BibitemOpen
  \bibfield  {author} {\bibinfo {author} {\bibfnamefont {S.}~\bibnamefont
  {Datta}},\ }\href@noop {} {\emph {\bibinfo {title} {Electronic transport in
  mesoscopic systems}}}\ (\bibinfo  {publisher} {Cambridge university press},\
  \bibinfo {year} {1997})\BibitemShut {NoStop}%
\bibitem [{\citenamefont {Altshuler}\ and\ \citenamefont
  {Aronov}(1985)}]{Altshuler85}%
  \BibitemOpen
  \bibfield  {author} {\bibinfo {author} {\bibfnamefont {B.~L.}\ \bibnamefont
  {Altshuler}}\ and\ \bibinfo {author} {\bibfnamefont {A.~G.}\ \bibnamefont
  {Aronov}},\ }in\ \href@noop {} {\emph {\bibinfo {booktitle} {Modern Problems
  in condensed matter sciences}}},\ Vol.~\bibinfo {volume} {10}\ (\bibinfo
  {publisher} {Elsevier},\ \bibinfo {year} {1985})\ pp.\ \bibinfo {pages}
  {1--153}\BibitemShut {NoStop}%
\bibitem [{\citenamefont {Li}\ \emph {et~al.}(2017)\citenamefont {Li},
  \citenamefont {Yu}, \citenamefont {Liu}, \citenamefont {Guan}, \citenamefont
  {Wang}, \citenamefont {Zhang}, \citenamefont {Yao},\ and\ \citenamefont
  {Yang}}]{Li17prb}%
  \BibitemOpen
  \bibfield  {author} {\bibinfo {author} {\bibfnamefont {S.}~\bibnamefont
  {Li}}, \bibinfo {author} {\bibfnamefont {Z.-M.}\ \bibnamefont {Yu}}, \bibinfo
  {author} {\bibfnamefont {Y.}~\bibnamefont {Liu}}, \bibinfo {author}
  {\bibfnamefont {S.}~\bibnamefont {Guan}}, \bibinfo {author} {\bibfnamefont
  {S.-S.}\ \bibnamefont {Wang}}, \bibinfo {author} {\bibfnamefont
  {X.}~\bibnamefont {Zhang}}, \bibinfo {author} {\bibfnamefont
  {Y.}~\bibnamefont {Yao}}, \ and\ \bibinfo {author} {\bibfnamefont {S.~A.}\
  \bibnamefont {Yang}},\ }\href {\doibase 10.1103/PhysRevB.96.081106}
  {\bibfield  {journal} {\bibinfo  {journal} {Phys. Rev. B}\ }\textbf {\bibinfo
  {volume} {96}},\ \bibinfo {pages} {081106} (\bibinfo {year}
  {2017})}\BibitemShut {NoStop}%
\bibitem [{\citenamefont {Ahn}\ \emph {et~al.}(2017)\citenamefont {Ahn},
  \citenamefont {Mele},\ and\ \citenamefont {Min}}]{Ahn17prl}%
  \BibitemOpen
  \bibfield  {author} {\bibinfo {author} {\bibfnamefont {S.}~\bibnamefont
  {Ahn}}, \bibinfo {author} {\bibfnamefont {E.~J.}\ \bibnamefont {Mele}}, \
  and\ \bibinfo {author} {\bibfnamefont {H.}~\bibnamefont {Min}},\ }\href
  {\doibase 10.1103/PhysRevLett.119.147402} {\bibfield  {journal} {\bibinfo
  {journal} {Phys. Rev. Lett.}\ }\textbf {\bibinfo {volume} {119}},\ \bibinfo
  {pages} {147402} (\bibinfo {year} {2017})}\BibitemShut {NoStop}%
\bibitem [{\citenamefont {Huang}\ \emph {et~al.}(2018)\citenamefont {Huang},
  \citenamefont {Jiang}, \citenamefont {Jin},\ and\ \citenamefont
  {Liu}}]{Huang18prb}%
  \BibitemOpen
  \bibfield  {author} {\bibinfo {author} {\bibfnamefont {H.}~\bibnamefont
  {Huang}}, \bibinfo {author} {\bibfnamefont {W.}~\bibnamefont {Jiang}},
  \bibinfo {author} {\bibfnamefont {K.-H.}\ \bibnamefont {Jin}}, \ and\
  \bibinfo {author} {\bibfnamefont {F.}~\bibnamefont {Liu}},\ }\href {\doibase
  10.1103/PhysRevB.98.045131} {\bibfield  {journal} {\bibinfo  {journal} {Phys.
  Rev. B}\ }\textbf {\bibinfo {volume} {98}},\ \bibinfo {pages} {045131}
  (\bibinfo {year} {2018})}\BibitemShut {NoStop}%
\bibitem [{\citenamefont {Rui}\ \emph {et~al.}(2018)\citenamefont {Rui},
  \citenamefont {Zhao},\ and\ \citenamefont {Schnyder}}]{Rui18prb}%
  \BibitemOpen
  \bibfield  {author} {\bibinfo {author} {\bibfnamefont {W.~B.}\ \bibnamefont
  {Rui}}, \bibinfo {author} {\bibfnamefont {Y.~X.}\ \bibnamefont {Zhao}}, \
  and\ \bibinfo {author} {\bibfnamefont {A.~P.}\ \bibnamefont {Schnyder}},\
  }\href {\doibase 10.1103/PhysRevB.97.161113} {\bibfield  {journal} {\bibinfo
  {journal} {Phys. Rev. B}\ }\textbf {\bibinfo {volume} {97}},\ \bibinfo
  {pages} {161113} (\bibinfo {year} {2018})}\BibitemShut {NoStop}%
\bibitem [{\citenamefont {Burkov}(2018)}]{Burkov18prb}%
  \BibitemOpen
  \bibfield  {author} {\bibinfo {author} {\bibfnamefont {A.~A.}\ \bibnamefont
  {Burkov}},\ }\href {\doibase 10.1103/PhysRevB.97.165104} {\bibfield
  {journal} {\bibinfo  {journal} {Phys. Rev. B}\ }\textbf {\bibinfo {volume}
  {97}},\ \bibinfo {pages} {165104} (\bibinfo {year} {2018})}\BibitemShut
  {NoStop}%
\bibitem [{\citenamefont {Son}\ and\ \citenamefont {Spivak}(2013)}]{Son13prb}%
  \BibitemOpen
  \bibfield  {author} {\bibinfo {author} {\bibfnamefont {D.~T.}\ \bibnamefont
  {Son}}\ and\ \bibinfo {author} {\bibfnamefont {B.~Z.}\ \bibnamefont
  {Spivak}},\ }\href {\doibase 10.1103/PhysRevB.88.104412} {\bibfield
  {journal} {\bibinfo  {journal} {Phys. Rev. B}\ }\textbf {\bibinfo {volume}
  {88}},\ \bibinfo {pages} {104412} (\bibinfo {year} {2013})}\BibitemShut
  {NoStop}%
\bibitem [{\citenamefont {Dai}\ \emph {et~al.}(2017)\citenamefont {Dai},
  \citenamefont {Du},\ and\ \citenamefont {Lu}}]{Dai17prl}%
  \BibitemOpen
  \bibfield  {author} {\bibinfo {author} {\bibfnamefont {X.}~\bibnamefont
  {Dai}}, \bibinfo {author} {\bibfnamefont {Z.~Z.}\ \bibnamefont {Du}}, \ and\
  \bibinfo {author} {\bibfnamefont {H.-Z.}\ \bibnamefont {Lu}},\ }\href
  {\doibase 10.1103/PhysRevLett.119.166601} {\bibfield  {journal} {\bibinfo
  {journal} {Phys. Rev. Lett.}\ }\textbf {\bibinfo {volume} {119}},\ \bibinfo
  {pages} {166601} (\bibinfo {year} {2017})}\BibitemShut {NoStop}%
\end{thebibliography}
\end{document}